\documentclass[structabstract]{aa}
\usepackage{graphicx}
\usepackage{color}
\usepackage{amsmath}
\usepackage{amssymb}
\usepackage{latexsym}
\usepackage{times}
\usepackage[varg]{txfonts}
\begin{document}
   \title{Kelvin-Helmholtz instability of kink waves in \\ photospheric twisted
   flux tubes}

   \author{I.~Zhelyazkov$^1$ and T.~V.~Zaqarashvili$^{2,3}$}

   \institute{$^1$ Faculty of Physics, Sofia University,
              5 James Bourchier Blvd., BG-1164 Sofia, Bulgaria\\
              \phantom{$^1$ }\email{izh@phys.uni-sofia.bg}\\
              $^2$ Space Research Institute, Austrian Academy of Sciences,
              Schmiedlstrasse 6, \\
              \phantom{$^2$ }A-8042 Graz, Austria\\
              \phantom{$^2$ }\email{teimuraz.zaqarashvili@oeaw.ac.at}\\
              $^3$ Abastumani Astrophysical Observatory at Ilia State University,
              2 University Street,\\
              \phantom{$^2$ }GE-0160 Tbilisi, Georgia}

   \date{Received August 11 2012, Accepted September 11 2012}


  \abstract
  {} 
   {We investigate conditions under which kink magnetohydrodynamic waves propagating along
   photospheric uniformly twisted flux tubes with axial mass flows become unstable
   as a consequence of the Kelvin-Helmholtz instability.}
   {We employed the dispersion relations of kink waves derived from the linearised
   magnetohydrodynamic equations.  We assumed real wave numbers and complex angular wave
   frequencies, namely complex wave phase velocities.  The dispersion relations were solved
   numerically at fixed input parameters and several mass flow velocities.}
   {We show that the stability of the waves depends upon four parameters, the
   density contrast between the flux tube and its environment, the ratio of
   the background magnetic fields in the two media, the twist of the magnetic
   field lines inside the tube, and the value of the Alfv\'en-Mach number (the ratio
   of the jet velocity to Alfv\'en speed inside the flux tube).  We assume that the
   azimuthal component of the magnetic field in the tube is proportional to
   the distance from the tube axis and that the tube is only weakly twisted (i.e.,
   the ratio of the azimuthal and axial components of the magnetic field is low).
   At certain densities and magnetic field twists, an instability of the
   Kelvin-Helmholtz type of kink ($m = 1$) mode can arise if the Alfv\'en-Mach number
   exceeds a critical value.  In particular, for an isolated
   twisted magnetic flux tube (magnetically free environment) at a density
   contrast (the ratio of the mass density of the environment to that of the tube
   itself) equal to $2$ and a magnetic field twist (defined as the ratio of azimuthal
   magnetic field component at the inner surface of the tube to the background
   magnetic field strength) equal to $0.4$, the threshold Alfv\'en-Mach number has a
   magnitude of $1.250075$, which means that for an Alfv\'en speed inside the tube of
   $10$~km\,s$^{-1}$ the jet velocity should be higher than $12.5$~km\,s$^{-1}$ to
   ensure the onset of the Kelvin-Helmholtz instability of the kink ($m = 1$) mode.
   Speeds of that order can be detected in photospheric tubes.}
   {The observed mass flows may trigger the Kelvin-Helmholtz instability of the kink ($m = 1$)
   mode in weakly twisted photospheric magnetic flux tubes at critical Alfv\'en-Mach
   numbers lower that those in untwisted tubes if the magnetic field twist lies in the
   range $0.36$--$0.4$ and the flow speed exceeds a critical value.  A weak
   external magnetic field (with a ratio to the magnetic field inside the tube
   in the range $0.1$--$0.5$) slightly increases that critical value.}


   \authorrunning{I.~Zhelyazkov \& T.~V.~Zaqarashvili}
   \titlerunning{Kelvin-Helmholtz instability of kink waves in twisted magnetic tubes}
   \maketitle
%

\section{Introduction}
\label{sec:intro}

With the wealth of recent high-resolution observations from SDO, Hinode, Stereo,
TRACE, SOHO, and RHESSI as well as from ground-based observations, the concepts of
the magnetic flux tube and coronal loops have become increasingly important for
understanding explosive phenomena such as solar flares and eruptive prominences
and also for the problem of the solar corona heating mechanism.  The observations
reveal a wide range of shapes and sizes, ranging from the small-size magnetic flux
tubes found in the granular network, sunspots, coronal loops, and arcades with many
loops to huge loops such as prominences.  The registered mass flows speeds in these
magnetic tubes range from a few tens of kilometers per second in the
photosphere/chromosphere (Shibata et al.\ \cite{shibata07}; Katsukawa et al.\
\cite{katsukawa07}; De Pontieu et al.\ \cite{bart07}) up to hundreds or thousands of
kilometers per second in coronal holes and X-ray jets (Shimojo \& Shibata
\cite{shimojo00}; Cirtain et al.\ \cite{cirtain07}; Ofman \& Wang \cite{ofman08};
Wallace et al.\ \cite{wallace10}; Madjarska \cite{madjarska11}; Kamio et al.\
\cite{kamio11}; Tian et al.\ \cite{tian11}; Srivastava \& Murawski
\cite{srivastava11}).

An additional feature of the flux tube is its twist.  There is observational
evidence for a twist in the solar atmosphere.  Aschwanden et al.\
(\cite{aschwanden99}) confirmed the existence of twisted loops in the corona by detecting a post-flare loop system in extreme ultraviolet wavelengths ($171$ \AA) with the Coronal Explorer (TRACE).  Rotational movement along a loop as
observed in the active region NOAA 8668 of the photosphere by Chae et al.\
(\cite{chae01}) indicates that there is a twist in the magnetic field lines.  More
recently, Srivastava et al.\ (\cite{srivastava10}), using multi-wavelength
observations of SOHO/MDI, SOT-Hinode/blue-continuum (\text{$4504$ \AA}), $G$ band
($4305$ \AA), \mbox{Ca\,\large \textsc{ii}} H ($3968$ \AA), and TRACE $171$ \AA,
presented the observational signature of a strongly twisted magnetic loop in AR
10960 during the period 04:43 UT--04:52 UT on June 4, 2007.

It is well-known that the hydromagnetic flows are generally unstable against the
Kelvin-Helmholtz instability when the flow speed exceeds a certain
threshold/critical value (Chandrasekhar \cite{chandra61}).  Recent observations of
Kelvin-Helmholtz vortices (Foullon et al.\ \cite{foullon11}; Ofman \& Thompson
\cite{ofman11}) increased the interest in flow instability in the solar corona.
On the other hand, the twisted magnetic field itself
can cause the so-called \emph{kink instability\/} even when there is no flow.  The
stability of static twisted magnetic flux tubes has also been studied extensively in
the context of laboratory plasma (e.g., Shafranov \cite{shafranov57}; Kruskal et al.\
\cite{kruskal58}).  Magnetic tubes are subject to the kink instability when the
twist exceeds a critical value (Lundquist \cite{lundquist51}; Hood \& Priest
\cite{hood79}).  Oscillations and waves and their stability in twisted magnetic
flux tubes without flow have been studied in the framework of the normal mode analysis
in earlier works (see Dungey \& Loughhead \cite{dungey54}; Roberts \cite{roberts56};
Trehan \cite{trehan58}; Bogdan \cite{bogdan84}; Linton et al.\ \cite{linton96};
Bennett et al. \cite{bennett99}; Erd\'{e}lyi \& Carter \cite{erdelyi06a};
Erd\'{e}lyi \& Fedun \cite{erdelyi06,erdelyi07,erdelyi10}; Ruderman
\cite{ruderman07}; Carter \& Erd\'{e}lyi \cite{carter08}; Ruderman \&
Erd\'{e}lyi \cite{ruderman09}; Karami \& Barin \cite{karami09}; and Karami \&
Bahari \cite{karami10}).  Most of these papers deal with relatively simple
twisted magnetic configurations: incompressible plasma cylinders/slabs surrounded
by perfectly conducting unmagnetised plasma or a medium with an untwisted
homogeneous magnetic field.  Erd\'{e}lyi \& Fedun (\cite{erdelyi06}) were the first
to study the wave propagation in a twisted cylindrical magnetic flux tube
embedded in an incompressible but also magnetically twisted plasma.  A more
complex magnetic field configuration has been studied by Erd\'{e}lyi \& Carter
(\cite{erdelyi06a}), who investigated the propagation of magnetohydrodynamic (MHD) waves in a fully magnetically twisted flux tube consisting of a core, an annulus, and an external region.
These authors considered magnetic twist just in the annulus, while the internal and external
regions retained a straight magnetic field.  Two modes of sausage oscillations occur
in this configuration, notably pure surface (i.e., evanescent) and hybrid
(spatially oscillatory in the twisted annulus, otherwise evanescent) waves.
The propagation characteristics of the kink mode in the same magnetically twisted
annulus have been studied by Carter \& Erd\'{e}lyi (\cite{carter08}).  The
influence of compressibility on the dispersion characteristics of the sausage mode
($m = 0$) propagating in a magnetically twisted flux tube embedded in a compressible
uniformly magnetised plasma environment in cylindrical geometry has been
investigated by Erd\'{e}lyi \& Fedun (\cite{erdelyi07}).  That analysis has been
generalised by the authors (see Erd\'{e}lyi \& Fedun \cite{erdelyi10}) for the
kink mode ($m = 1$), as well as for the fluting modes ($m > 1$); $m$ being the
azimuthal order of a given mode.  The last work by Erd\'{e}lyi \& Fedun
(\cite{erdelyi10}) -- the most elaborate study to date -- deals with the MHD
wave propagation along twisted
magnetic tubes of compressible ideal plasma surrounded by also compressible
but uniformly magnetised fully ionized medium.  The effect of a twisted
magnetic field on the spectra and the resonant absorption of kink and fluting
surface modes in resistive incompressible cylindrical plasmas have been studied
by Karami \& Barin (\cite{karami09}) and Karami \& Bahari (\cite{karami10}).
It was found that a magnetic twist will increase the frequencies, damping rates,
and the ratio of the oscillation frequency to the damping rate of these modes.
The period ratio $P_1/P_2$ of the fundamental and its first-overtone surface
waves for kink ($m = 1$) and fluting ($m = 2,3$) modes is lower than two (the
value for an untwisted loop) in the presence of a twisted magnetic field.  In
particular, for the kink modes, the magnetic twists $B_{\phi}/B_z = 0.0065$
and $0.0255$ can achieve deviations from two of the same order of magnitude
as in the observations.  The only work studying the wave propagation in a
twisted magnetic tube with a mass density variation along the tube is that of
Ruderman (\cite{ruderman07}).  He investigated non-axisymmetric oscillations
of thin (tube radius $a$ much smaller than the finite tube length $L$)
twisted magnetic tubes in a zero-beta plasma.  With an asymptotic analysis,
Ruderman (\cite{ruderman07}) showed that the eigenmodes and the
eigenfrequencies of the kink and fluting oscillations are described by a
classical Sturm-Liouville problem for a second-order ordinary differential
equation.  He also concluded that the results concerning non-axisymmetric
waves in twisted magnetic tubes obtained by Bennett et al.\
(\cite{bennett99}) for incompressible plasmas can be applied to global
non-axisymmetric modes, i.e., kink and fluting modes, in coronal loop
even though the coronal plasma is a low-beta plasma.  This conclusion 
also agrees well with the results obtained by Erd\'{e}lyi \& Fedun
(\cite{erdelyi07}).

An open question is how a flow along a twisted magnetic flux
tube will change the dispersion properties of the propagating modes and
their stability.  It turns out that the flow may decrease the threshold
for the kink instability, as was tested experimentally in a laboratory
twisted plasma column (Furno et al.\ \cite{furno07}).  This observation was
theoretically confirmed by Zaqarashvili et al.\ (\cite{temury10}).  The
authors studied the influence of axial mass flows on the stability of an
isolated twisted magnetic tube of incompressible plasma embedded in
a perfectly conducting unmagnetised plasma.  Two main results were
found.  First, the axial mass flow reduces the threshold of the kink
instability in twisted magnetic tubes.  Second, the twist of the
magnetic field leads to the Kelvin-Helmholtz instability of
sub-Alfv\'enic flows for the harmonics with a sufficiently large azimuthal
mode number $m$.  The effect is more significant for photospheric
magnetic flux tubes than for coronal ones.  In a recent paper,
D\'{i}az et al.\ (\cite{diaz11}) also studied the equilibrium and
stability of twisted magnetic flux tubes with mass flows, but
for flows along the field lines.  The authors focused on the stability
and oscillatory modes of magnetic tubes with a uniform twist in a
zero-beta plasma surrounded by a uniform cold plasma embedded in a
purely longitudinal magnetic filed.  Regarding the equilibrium, the
authors claimed that the only value of the flow that satisfies the
equations for their magnetic field configuration is a super-Alfv\'enic
one.  The main conclusion is that the twisted tube is subject to the
kink instability unless the magnetic field pitch is very high, since
the Lundquist criterion is significantly lowered.  This is caused by
the requirement of having an Alfv\'en-Mach number greater than $1$,
so the magnetic pressure balances the magnetic field tension and
fluid inertia.  The authors suggest that this type of instability might
be observed in some solar atmospheric structures, such as surges.

Kink waves are frequently observed in the chromospheric spectral lines
as a periodic Doppler shift and/or periodic spatial displacement of the tube
axis (Kukhianidze et al.\ \cite{kuk06}; Zaqarashvili et al.\
\cite{temury07}; Zaqarashvili \& Erd\'{e}lyi \cite{temury09}; He et al.\
\cite{he09}; Pietarila et al. \cite{pietarila11}; Tavabi et al.\
\cite{tavabi11}; Ebadi et al.\ \cite{ebadi12}).  Most of the observations
have been associated to spicules, which are chromospheric plasma jets.
It is well known that a simple hydrodynamic jet is unstable against the
antisymmetric (kink) mode (Drazin \cite{drazin02}; Zaqarashvili
\cite{temury11}), therefore the observed periodic displacements can be
connected to the unstable kink mode of a plasma jet in spicules.  However,
the magnetic field of spicules may significantly influence the stability
of the kink mode in the plasma jet, therefore studying the instability of
$m = 1$ mode of the jet in a magnetic cylinder is of vital importance.

In the present work, our aim is to investigate the effect of a twisted
magnetic field on the stability of kink ($m = 1$) waves propagating on
a cylindrical incompressible radially homogeneous tube with axial
mass flow, assuming that the waves are subject, under certain conditions, to
the Kelvin-Helmholtz instability.  To evaluate that effect, we compare
the critical flow speeds at which the instability occurs with those in the
untwisted magnetic tubes.  It is necessary to point out that our
simplified model of radially homogeneous plasma density inside the tube
rules out the resonant absorption of MHD modes that can modify both the
wave frequency spectrum and conditions under which the Kelvin-Helmholtz
instability occurs.  The radial density inhomogeneity is usually introduced
as a nonuniform transition layer that continuously connects the internal
density to the external one.  Owing to this transverse
inhomogeneous transitional layer, wave modes with $m \neq 1$ are spatially
damped by resonant absorption.  In particular, the effect of longitudinal
flow on the resonantly damped propagating kink waves has been studied by
Terradas et al.\ (\cite{terradas10a,terradas10b}) and by Soler et al.\
(\cite{soler10a}).  We note that a Kelvin-Helmholtz instability of kink
oscillations in flux tubes with both a sharp and smooth transition layer can
also occur as a result of shear motions (Terradas et al.\ \cite{terradas08}).
But as Terradas et al.\ (\cite{terradas08}) claim, a magnetic twist, not
included in their model, might decrease or even suppress the instability
because a magnetic field component along the flow stabilises
the Kelvin-Helmholtz instability.  The same conclusion for the stabilising
role of a small azimuthal component of the magnetic field has been drawn in the
paper by Soler et al.\ (\cite{soler10b}).  The influence of a twisted magnetic
field on the stability status of kink waves propagating along radially
inhomogeneous solar cylindrical photospheric/chromospheric jets is beyond
the scope of this study, and as was mentioned above, we limit ourselves to
considering radially homogeneous flowing plasmas only.

This paper is organized as follows.  In Sec.~2 we
specify the geometry of the problem, governing equations, and the
derivation of the wave dispersion relation.  Section 3 deals with the
numerical solutions to the dispersion relation for various values of the input parameters and compares derived dispersion curves and growth rates of the
unstable kink waves with those in untwisted magnetic tubes.  The last
Sec.~4 summarises our results.

\section{Geometry, the basic MHD equations, and the wave dispersion relation}
\label{sec:basic}

We consider a magnetic flux tube with radius $a$ and density $\rho_{\rm i}$
embedded in a uniform field environment with density $\rho_{\rm e}$.  Both
media (inside and outside the tube) are supposed to be incompressible.  The
magnetic field inside the tube is helicoidal, $\vec{B}_{\rm i} = (0,
B_{{\rm i} \phi}(r), B_{{\rm i} z}(r))$, while outside the magnetic field
is uniform and directed along the $z$-axis, $\vec{B}_{\rm e} = (0, 0,
B_{\rm e})$.  Both $\rho_{\rm i}$ and $\rho_{\rm e}$ are assumed to
be homogeneous.  We consider the mass flow $\vec{v}_0 = (0,0,v_0)$ directed
along the $z$-axis; thus the equilibrium mass flow is not field-aligned.  No
mass flow is present outside the tube, therefore the surrounding photospheric
medium is considered to be uniformly magnetised ($B_{\rm e}\hat{z}$ is
constant), uniform ($\rho_{\rm e}$ constant), and lacking mass flow at
equilibrium.

In cylindrical equilibrium the magnetic field and plasma pressure satisfy
the equilibrium condition in the radial direction
\begin{equation}
\label{eq:equilib}
    \frac{\mathrm{d}}{\mathrm{d}r}\left( p_{\rm i} + \frac{B_{\rm i}^2}{2\mu}
    \right) = -\frac{B_{{\rm i} \phi}^2}{\mu r}.
\end{equation}
Here, $B_{\rm i}(r) = \left( B_{{\rm i} \phi}^2 + B_{{\rm i} z}^2 \right)^{1/2}
= \left| \vec{B}_{\rm i} \right|$ denotes the strength of the equilibrium
magnetic field, and $\mu$ is the magnetic permeability.  We note that in
Eq.~(\ref{eq:equilib}) the total (plasma plus magnetic) pressure gradient is
balanced by the tension force (the right-hand side of Eq.~(\ref{eq:equilib}))
in the twisted field.  We consider the special case of an equilibrium with
uniform twist, i.e., the one for which $B_{{\rm i} \phi}(r)/r B_{{\rm i} z}(r)$
is a constant.  Accordingly, the background magnetic field is assumed to be
\begin{equation}
\label{eq:magnfield}
    \vec{B}(r) = \left\{ \begin{array}{lc}
                     (0, Ar, B_{{\rm i} z}) & \mbox{for $r \leqslant a$}, \\
                     (0, 0, B_{\rm e}) & \mbox{for $r > a$},
                              \end{array}
                      \right.
\end{equation}
where $A$, $B_{{\rm i} z}$, and $B_{\rm e}$ are constant.  Then the equilibrium
condition (\ref{eq:equilib}) gives the equilibrium plasma pressure $p_{\rm i}(r)$
as
\[
    p_{\rm i}(r) = p_0 - \frac{A^2 r^2}{\mu},
\]
where $p_0$ is the plasma pressure at the centre of the tube.  We note that the
jump in the value of $B_{\phi}(r)$ across $r = a$ implies a surface current
there.

The plasma motion is governed by the set of linearised MHD equations for an ideal
incompressible plasma:
\begin{eqnarray}
\label{eq:moment}
	\rho \frac{\partial}{\partial t}\vec{v}_1 + \rho \left(
        \vec{v}_0 \cdot \nabla \right) \vec{v}_1 + \nabla \left( p_1
        + \frac{\vec{B}_0 \cdot \vec{B}_1}{\mu} \right) \nonumber \\
        \nonumber \\
        {}-\frac{1}{\mu} ( \vec{B}_0 \cdot \nabla ) \vec{B}_1 -
        \frac{1}{\mu} ( \vec{B}_1 \cdot \nabla ) \vec{B}_0 = 0,
\end{eqnarray}
\begin{equation}
\label{eq:induct}
	\frac{\partial}{\partial t}\vec{B}_1 - \nabla \times \left( \vec{v}_1
    \times \vec{B}_0 \right) - \nabla \times \left( \vec{v}_0 \times \vec{B}_1
    \right) = 0,	
\end{equation}
\begin{equation}
\label{eq:divv}
	\nabla \cdot \vec{v}_1 = 0,	
\end{equation}
\begin{equation}
\label{eq:divb}
	\nabla \cdot \vec{B}_1 = 0.
\end{equation}
Here, the index `$0$' denotes equilibrium values of the fluid velocity and the
medium magnetic field, and the index `$1$' their perturbations.  Below, the
sum $p_1 + \vec{B}_0 \cdot \vec{B}_1/\mu$ in Eq.~(\ref{eq:moment}) will
be replaced by $p_{\rm 1 tot}$, which represents the total pressure perturbation.

Assuming that all perturbations are proportional to $g(r)\exp \left[
\mathrm{i} \left( -\omega t + m \phi + k_z z \right) \right]$ with
$g(r)$ being just a function of $r$, and that in cylindrical coordinates the
nabla operator has the form
\[
	\nabla \equiv \frac{\partial}{\partial r}\hat{r} + \frac{1}{r}
        \frac{\partial}{\partial \phi}\hat{\phi} + \frac{\partial}{\partial
        z}\hat{z},
\]
from the above set of equations one can obtain a second-order differential
equation for the total pressure perturbation $p_{\rm 1 tot}$
\begin{equation}
\label{eq:diffeq}
	\left[ \frac{\mathrm{d}^2}{\mathrm{d}r^2} + \frac{1}{r}
        \frac{\mathrm{d}}{\mathrm{d} r} - \left( \kappa^2 + \frac{m^2}{r^2}
        \right) \right] p_{\rm 1 tot} = 0,
\end{equation}
as well as an expression for the radial component $v_{1r}$ of the fluid
velocity perturbation $\vec{v}_1$ in terms of $p_{\rm 1 tot}$ and its
first derivative
\begin{equation}
\label{eq:v1r}
	{v}_{1r} = -\mathrm{i}\frac{1}{\rho} \frac{1}{Y} \frac{\omega -
        \vec{k} \cdot \vec{v}_0}{\left( \omega - \vec{k} \cdot
        \vec{v}_0 \right)^2 - \omega_{\rm A}^2} \left(
        \frac{\mathrm{d}}{\mathrm{d} r} p_{\rm 1 tot} - Z \frac{m}{r}
        p_{\rm 1 tot} \right).
\end{equation}
In Eq.~(\ref{eq:diffeq}), $\kappa$ is the so-called \emph{wave attenuation
coefficient}, which characterises the space structure of the wave and whose
squared magnitude is given by the expression
\begin{equation}
\label{eq:kappa}
	\kappa^2 = k_z^2 \left(  1 - \frac{4 A^2 \omega_{\rm A}^2}
        {\mu \rho \left[ \left( \omega - \vec{k} \cdot
        \vec{v}_0 \right)^2 - \omega_{\rm A}^2\right]^2} \right),
\end{equation}
where
\begin{equation}
\label{eq:alfvenfrq}
	\omega_{\rm A} = \frac{1}{\sqrt{\mu \rho}}\left( mA + k_z B_z \right)
\end{equation}
is the so-called \emph{local Alfv\'en frequency\/} (Bennett et al.\
\cite{bennett99}).  The numerical coefficients $Z$ and $Y$ in the
expression of $v_{1r}$ (see Eq.~(\ref{eq:v1r})) are respectively
\[
    Z = \frac{2A \omega_{\rm A}}{\sqrt{\mu \rho}\left[ \left( \omega -
        \vec{k} \cdot \vec{v}_0 \right)^2 - \omega_{\rm A}^2\right]}
        \qquad \mbox{and} \qquad Y = 1 - Z^2.
\]

As seen from the expressions for the wave attenuation coefficient and the
radial fluid velocity component perturbation, the two quantities have different
values inside and outside the twisted flux tube owing to the different spatial
structure of the magnetic field in both media, and to the different mass
densities ($\rho_{\rm i}$ and $\rho_{\rm e}$, respectively).  It is important
to note that the attenuation coefficient $\kappa_{\rm e}$ outside the tube is
simply equal to $k_z$, whilst that inside the tube can become a purely imaginary
quantity, thus allowing the propagation of pseudosurface (body) waves along
the tube.  Notably the twisted magnetic field inside the tube enables the existence 
of these waves, which otherwise are pure surface modes
(Bennett et al.\ \cite{bennett99}).

Equation~(\ref{eq:diffeq}) has two different solutions inside and
outside the jet
\[
	p_{\rm 1 tot}(r) = \left\{ \begin{array}{lc}
         \alpha_{\rm i} I_m(\kappa_{\rm i} r) & \mbox{for $r \leqslant a$}, \\
         \alpha_{\rm e} K_m(k_z r) & \mbox{for $r > a$}.
                              \end{array}
                      \right.
\]
Here $I_m$ and $K_m$ are the modified Bessel functions of the first
and second kind, and $\alpha_{\rm i}$ and $\alpha_{\rm e}$ are arbitrary
constants.

The boundary conditions have to ensure that the normal component of the
interface perturbation
\[
    \xi_r = -\frac{v_{1r}}{\mathrm{i}\left( \omega - \vec{k} \cdot
        \vec{v}_0 \right)}
\]
remains continuous across the unperturbed tube boundary $r = a$, and also that
the total Lagrangian pressure is conserved across the perturbed boundary.  This
leads to the conditions (Bennett et al.\ \cite{bennett99})
\begin{equation}
\label{eq:contxi}
    \left[ \xi_r \right]^{\rm inner} = \left[ \xi_r \right]^{\rm outer},
\end{equation}
\begin{equation}
\label{eq:contp1tot}
    \left[ p_{\rm 1 tot} - \frac{B_{\phi}^2}{\mu a}\xi_r \right]^{\rm inner} =
    \left[ p_{\rm 1 tot} - \frac{B_{\phi}^2}{\mu a}\xi_r \right]^{\rm outer}.
\end{equation}

Applying boundary conditions (\ref{eq:contxi}) and (\ref{eq:contp1tot}) to our
solutions of $p_{\rm 1 tot}$ and $v_{1r}$ (and respectively $\xi_r$), we obtain 
after some algebra the dispersion relation of the normal modes propagating
along a twisted magnetic tube with axial mass flow $\vec{v}_0$
\begin{eqnarray}
\label{eq:dispeq}
	\frac{\left[ \left( \omega - \vec{k} \cdot \vec{v}_0 \right)^2 -
    \omega_{\rm Ai}^2 \right]{\displaystyle \frac{\kappa_{\rm i}a I_m^{\prime}(\kappa_{\rm
    i}a)}{I_m(\kappa_{\rm i}a)}} - 2m \omega_{\rm Ai}
    {\displaystyle \frac{A}{\sqrt{\mu \rho_{\rm i}}}}}
    {\left[ \left( \omega - \vec{k} \cdot \vec{v}_0 \right)^2 -
    \omega_{\rm Ai}^2 \right]^2 - 4\omega_{\rm Ai}^2
    {\displaystyle \frac{A^2}{\mu \rho_{\rm i}}} } \nonumber \\
    \nonumber \\
    {}= \frac{{\displaystyle \frac{k_z a K_m^{\prime}(k_z a)}{K_m(k_z a)}}}
    {{\displaystyle \frac{\rho_{\rm e}}{\rho_{\rm i}}} \left( \omega^2 - \omega_{\rm Ae}^2
    \right) + {\displaystyle \frac{A^2}{\mu \rho_{\rm i}}
    \frac{k_z a K_m^{\prime}(k_z a)}{K_m(k_z a)}}}.
\end{eqnarray}

This dispersion relation is a generalisation of Eq.~(23) in the work by Bennett et al.\
(\cite{bennett99}) that is applicable to a twisted magnetic flux tube without flow and of Eq.~(13) in the work by Zaqarashvili et al.\ (\cite{temury10}) that is applicable to a twisted magnetic flux
tube with a flow embedded in a non-magnetic environment.  As we have already seen, the wave
frequency $\omega$ is Doppler-shifted inside the jet.  In the above equation a prime ($\prime$)
denotes the derivative of a Bessel function to its dimensionless argument, and the local
Alfv\'en frequencies inside and outside the tube are correspondingly
\[
    \omega_{\rm Ai} = \frac{1}{\sqrt{\mu \rho_{\rm i}}}\left( mA + k_z B_{{\rm i}z}
    \right) \qquad \mbox{and} \qquad \omega_{\rm Ae} = \frac{k_z B_{\rm e}}{\sqrt{\mu
    \rho_{\rm e}}}.
\]
The wave attenuation coefficient inside the tube, according to
Eq.~(\ref{eq:kappa}), is given by
\[
    \kappa_{\rm i} = k_z \left(  1 - \frac{4 A^2 \omega_{\rm Ai}^2}
        {\mu \rho_{\rm i} \left[ \left( \omega - \vec{k} \cdot
        \vec{v}_0 \right)^2 - \omega_{\rm Ai}^2\right]^2} \right)^{1/2}.
\]

Assuming no twist, when $B_{{\rm i} \phi} = 0$, the dispersion relation
(\ref{eq:dispeq}) reduces to
\begin{eqnarray}
\label{eq:edwineq}
    \frac{\rho_{\rm e}}{\rho_{\rm i}}\left( \omega^2 - k_z^2 v_{\rm Ae}^2 \right)
    \frac{I_m^{\prime}(k_z a)}{I_m(k_z a)} \nonumber \\
    \nonumber \\
    {-} \left[ \left( \omega - \vec{k}
    \cdot \vec{v}_0 \right)^2 - k_z^2 v_{\rm Ai}^2 \right]
    \frac{K_m^{\prime}(k_z a)}{K_m(k_z a)} = 0.
\end{eqnarray}
The Alfv\'en speeds in this equation are expressed in terms of the background
magnetic fields in both media and the corresponding mass densities via the
well-known definitions
\[
    v_{\rm Ai} = \frac{B_{\rm i}}{\sqrt{\mu \rho_{\rm i}}} \qquad \mbox{and}
    \qquad v_{\rm Ae} = \frac{B_{\rm e}}{\sqrt{\mu \rho_{\rm e}}},
\]
respectively.  Equation (\ref{eq:edwineq}) is akin to a dispersion relation in
the same form as was previously obtained by Edwin \& Roberts (\cite{pat83})
for untwisted magnetic tubes without flow.  We also recall that for the
kink mode ($m = 1$) one defines the so-called \emph{kink speed\/} (Edwin \&
Roberts \cite{pat83})
\begin{equation}
\label{eq:kinkspeed}
	c_{\rm k} = \left( \frac{\rho_{\rm i} v_{\rm Ai}^2 + \rho_{\rm e}
        v_{\rm Ae}^2}{\rho_{\rm i} + \rho_{\rm e}} \right)^{1/2} = \left(
        \frac{v_{\rm Ai}^2 + (\rho_{\rm e}/\rho_{\rm i})v_{\rm Ae}^2}{1 +
        \rho_{\rm e}/\rho_{\rm i}} \right)^{1/2},
\end{equation}
which characterises the propagation of transverse perturbations.

\section{Numerical solutions and wave dispersion diagrams}
\label{sec:dispers}

We focus our study on the propagation of the kink mode, i.e., for $m = 1$.
It is obvious that Eq.~(\ref{eq:dispeq}) and Eq.~(\ref{eq:edwineq}) can be
solved only numerically.  The first step is to define the input parameters
that characterise the twisted magnetic tube and the axial flow, and to
normalise all variables in the dispersion equations.  The density contrast
between the tube and its environment is characterised by the parameter
$\eta = \rho_{\rm e}/\rho_{\rm i}$.  In investigating the kink instability in
twisted magnetic tubes with axial and field-aligned flows with
their equilibrium magnetic fields in the form $\vec{B}(r) = (0, Ar, B_0)$,
Zaqarashvili et al.\ (\cite{temury10}) and D\'{i}az et al.\ (\cite{diaz11})
chose as twist characteristics the so-called `dimensionless pitch' $k_z p$,
in which the specific length $p$ is defined (in their notation) as
\[
    p = \frac{B_0}{A} \qquad \mbox{or more generally as} \qquad
    p = \frac{B_z}{A}.
\]
That dimensionless pitch can be presented in the form $k_z p = 2 \pi p/\lambda$,
where $2 \pi p$ is the magnetic field pitch in its common sense.  As we showed, this
parameter is associated with a fixed wavelength $\lambda$.  To study the
Kelvin-Helmholtz instability in these tubes we prefer to specify the twist by
the ratio of the two magnetic field components, $B_{\phi}$ and $B_z$, evaluated
at the inner boundary of the tube, $r = a$, i.e., via $B_{\phi}/B_z$, where
$B_{\phi} = Aa$.  This choice is symbolically equivalent to the replacement of
the wavelength $\lambda$ by the tube radius $a$.  Following Aschwanden
(\cite{aschwanden05}), the magnetic twist parameter $B_{\phi}/B_z$ can be
estimated as $B_{\phi}/B_z = \tan \theta$, where $\theta$ is the shear angle
between the untwisted and twisted field line.  The geometric shear angle,
$\theta$, can observationally be measured in twisted photospheric magnetic
tubes and in twisted coronal loops.  We note that our choice of the twist
parameter allows us to immediately compare the dispersion curves of kink
waves and the growth rates when the waves become unstable in twisted
magnetic tubes with those in untwisted ones.

As usual, we normalise the velocities to the Alfv\'en speed $v_{\rm Ai} =
B_{\rm i}/(\mu \rho_{\rm i})^{1/2}$, where $B_{\rm i} = \left(
B_{{\rm i}\phi}^2 + B_{{\rm i}z}^2 \right)^{1/2}$ is evaluated at the inner
boundary of the tube, $r = a$.  Thus, we introduce the dimensionless wave
phase velocity $V_{\rm ph} = v_{\rm ph}/v_{\rm Ai}$ and the
\emph{Alfv\'en-Mach number\/} $M_{\rm A} = v_0/v_{\rm Ai}$, the latter
characterising the axial mass flow.  The wavelength is normalised to the
tube radius $a$, which means that the dimensionless wave number is
$K = k_z a$.  It is worth to discuss the normalisation of the local Alfv\'en
frequency $\omega_{\rm Ai}$.  After multiplying both sides of
Eq.~(\ref{eq:alfvenfrq}) by the tube radius $a$ and extracting a $B_{\rm i}$
from the brackets, we derive that
\[
    a \omega_{\rm Ai} = \frac{B_{\rm i}}{\sqrt{\mu \rho_{\rm i}}} \left(
    m \frac{B_{{\rm i}\phi}}{B_{\rm i}} + k_z a \frac{B_{{\rm i}z}}{B_{\rm i}}
    \right), \quad \mbox{where} \quad
    \frac{B_{{\rm i}z}}{B_{\rm i}} = \left( 1 - \frac{B_{{\rm i}\phi}^2}
    {B_{\rm i}^2} \right)^{1/2}.
\]
To simplify the numerical code, and largely for clarity, we can redefine
our twist parameter as $B_{{\rm i}\phi}/B_{\rm i} \equiv \varepsilon$, and consequently
the normalised local Alfv\'en frequency is expressed in terms of the Alfv\'en
speed $v_{\rm Ai}$, the dimensionless wave number $K$, and $\varepsilon$:
\[
    a \omega_{\rm Ai} = v_{\rm Ai}\left( m \varepsilon + K b_z \right),
\]
where the dimensionless axial magnetic field component $b_z = \left( 1 -
\varepsilon^2 \right)^{1/2}$.  The new twist parameter, $\varepsilon$, is
related to the real twist $B_{\phi}/B_{z}$ via the simple relation
\begin{equation}
\label{eq:epsilon}
    \varepsilon = \frac{B_{\phi}/B_{z}}{\sqrt{1 + \left( B_{\phi}/B_{z}
    \right)^2}}.
\end{equation}
Thus, if we have an observationally measured shear angle $\theta$ of some
solar twisted magnetic tube, then $B_{\phi}/B_{z} = \tan \theta$ can be
inserted into expression~(\ref{eq:epsilon}) to obtain the adopted twist
$\varepsilon$ of that specific magnetic tube.

Before starting the numerical calculation, we have to specify the values of the input
parameters.  First we will examine, as in the work of
Zaqarashvili et al.\ (\cite{temury10}), the case of an isolated twisted magnetic
tube with axial mass flow, i.e., with $B_{\rm e} = 0$, and hence for
parameter $b \equiv B_{\rm e}/B_{\rm i} = 0$.  Then, we will study the
influence of the external magnetic field on the instability onset by choosing $b$
equal to $0.1$ and $0.5$, respectively.  Concerning the density contrast,
$\eta$, we consider two cases, namely $\eta = 2$ and $\eta = 0.1$.  Following
Ruderman (\cite{ruderman07}), to satisfy the Shafranov-Kruskal stability
criterion for a kink instability, we assume that the azimuthal component of the
magnetic field is smaller than its axial component, i.e., we will choose our
twist parameter $\varepsilon$ to be always less than $1$.  In particular, we will
study the dispersion diagrams of kink waves and their growth rates (when the
waves are unstable) for three fixed values of $\varepsilon$: $0$ (untwisted
magnetic tube), $0.1$, and $0.4$.  The Alfv\'en-Mach number $M_{\rm A}$ will
take values from zero (no flow) to any reasonable number.

\subsection{Kink waves in twisted magnetic tubes with a density contrast
            $\eta = 2$}
\label{subsec:eta2}

\begin{figure}[ht]
   \centering
   \includegraphics[height=.23\textheight]{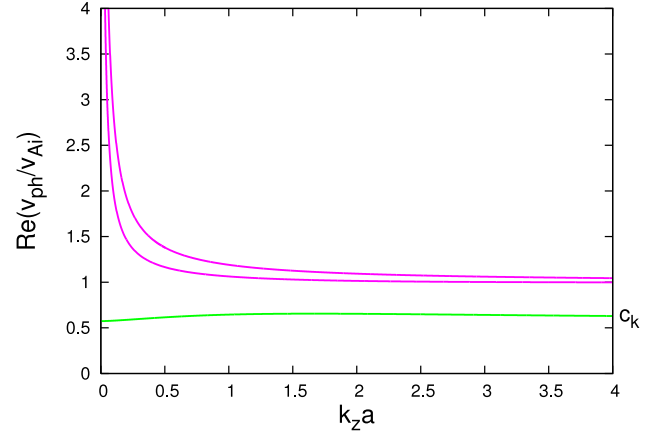}
   \caption{Dispersion curves of kink waves propagating along a twisted magnetic
            flux tube at $\eta = 2$, $B_{\rm e}/B_{\rm i} = 0$,
            $B_{{\rm i}\phi}/B_{\rm i} = 0.1$, and
            $M_{\rm A} = 0$.  The green curve is the dispersion curve of the wave
            associated with the kink speed $c_{\rm k}$.  The two purple curves
            depict a band that contains nine dispersion curves of similar
            form, not plotted here.}
   \label{fig:fig1}
\end{figure}

We start by calculating the dispersion curves of kink waves assuming that the
twist parameter $\varepsilon = 0.1$, and that the angular wave frequency
$\omega$ is real.  As a reference, we first assume that the plasma in the flux
tube is static, i.e., $M_{\rm A} = 0$.  The dispersion curves, which present
the dependence of the normalised wave phase velocity, $v_{\rm ph}/v_{\rm Ai}$,
on the normalised wave number, $k_z a$, are in this case shown in
Fig.~\ref{fig:fig1}.  One can recognise two types of waves: a sub-Alfv\'enic
wave labelled \textsf{c}$_{\mathsf{k}}$ (the green curve), and a family of
super-Alfv\'enic waves accommodated in a narrow band depicted by the two
purple dispersion curves.  The green dispersion curve is related to the
kink speed defined by Eq.~(\ref{eq:kinkspeed}).
For a weakly twisted magnetic tube ($\varepsilon = 0.1$) with a density
contrast $\eta = 2$ the kink speed is equal to $0.5774 v_{\rm Ai}$ -- our
numerical code yields in the long wavelength limit,
$c_{\rm k} = 0.5738 v_{\rm Ai}$, which agrees well with the value
calculated from Eq.~(\ref{eq:kinkspeed}).  If we assume that the Alfv\'en
speed inside the tube is typically $10$ km\,s$^{-1}$, then the kink speed
in the twisted magnetic tube is equal to $5.7$ km\,s$^{-1}$, or
approximately to $6$ kilometers per second.  We recall that the
expression of $c_{\rm k}$ was defined for untwisted magnetic tubes and
there is no guarantee that it will be valid for twisted tubes as well.
\begin{figure*}[ht]
  \centering
  \begin{tabular}{ccc}
    \includegraphics[width=6.5cm]{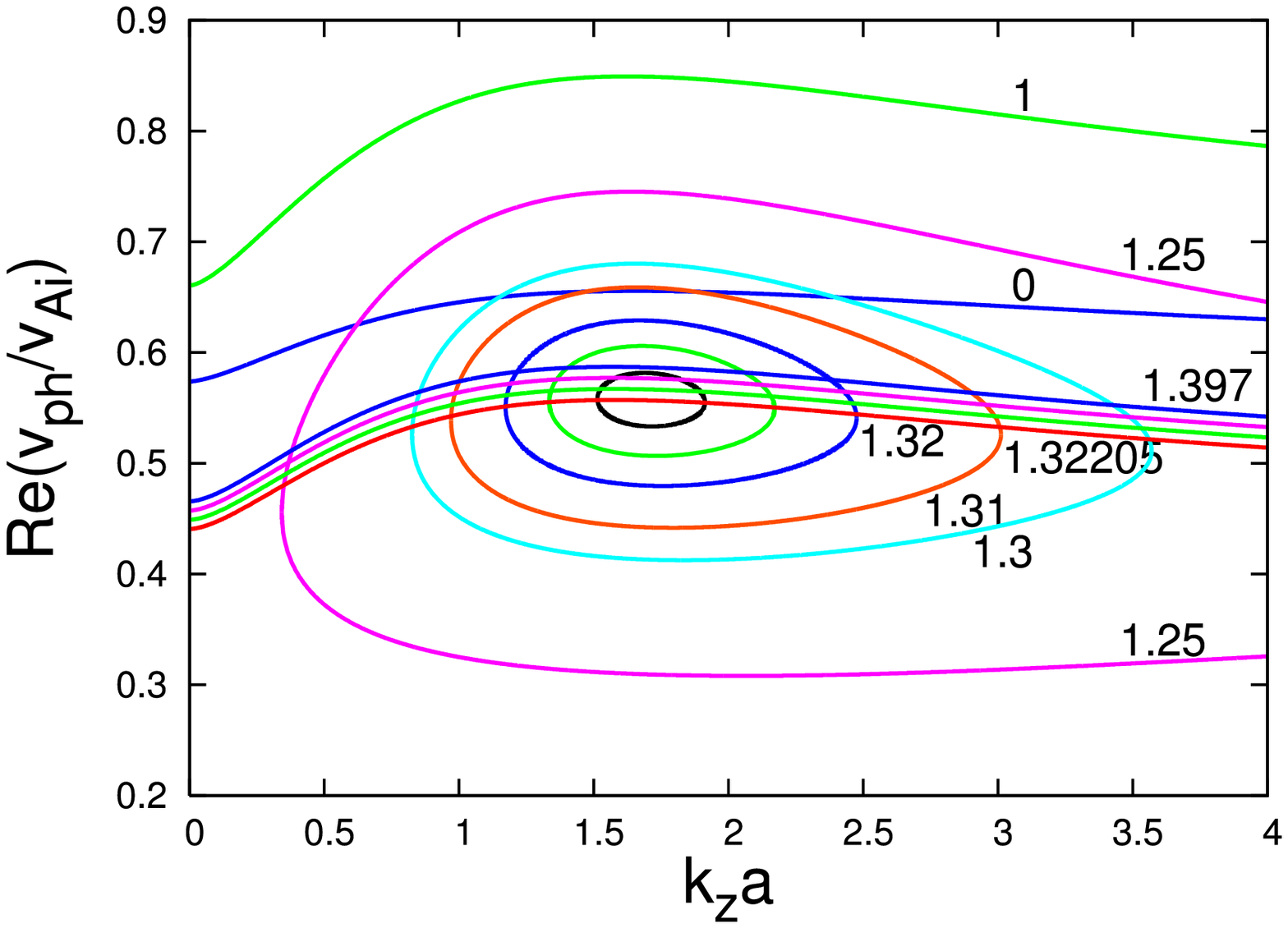} &
    \hspace{10mm} &
    \includegraphics[width=6.5cm]{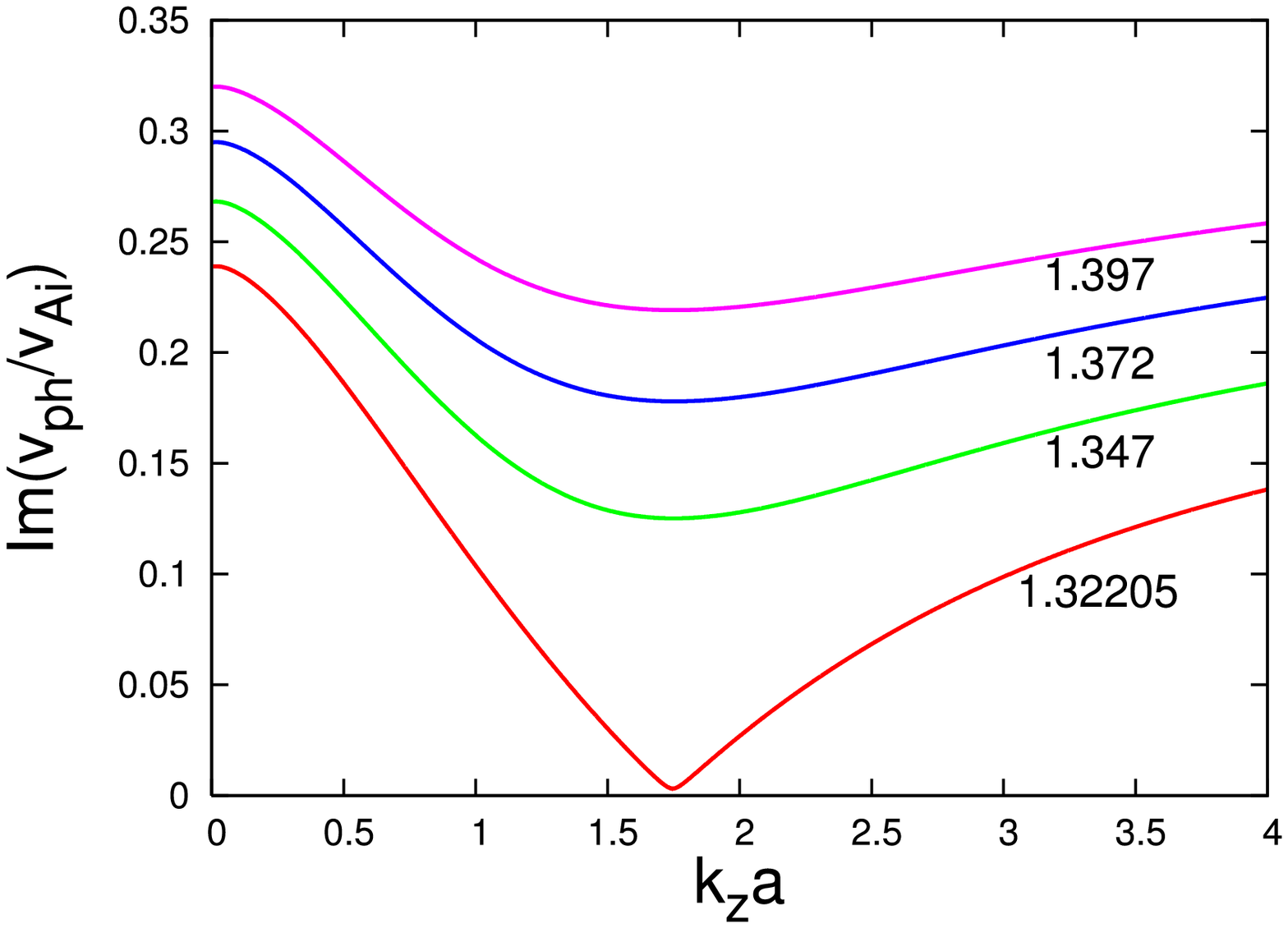}
  \end{tabular}
  \caption{(\emph{Left panel}) Dispersion curves of forward propagating stable
  and unstable kink waves in a twisted magnetic flux tube with the same input
  parameters as in Fig.~\ref{fig:fig1} for various values of the Alfv\'en-Mach
  number $M_{\rm A}$.  The non-labelled green closed curve corresponds to
  $M_{\rm A} =  1.325$, and the black one -- to $M_{\rm A} = 1.328$.
  (\emph{Right panel}) Growth rates of the unstable kink waves for
  $M_{\rm A} = 1.32205, 1.347, 1.372$, and $1.397$.}
  \label{fig:fig2}
\end{figure*}
\begin{figure*}[ht]
  \centering
  \begin{tabular}{ccc}
    \includegraphics[width=6.5cm]{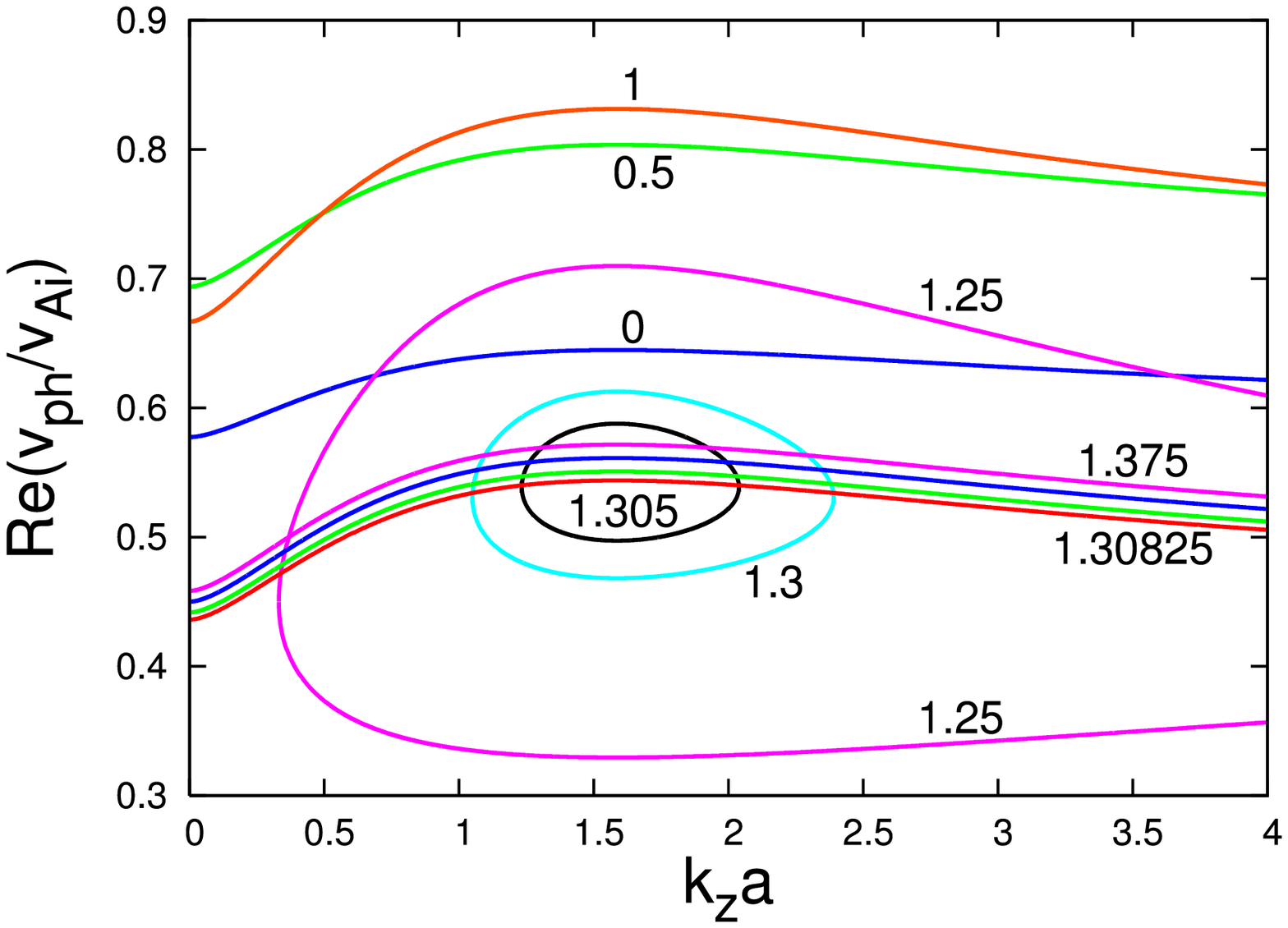} &
    \hspace{10mm} &
    \includegraphics[width=6.5cm]{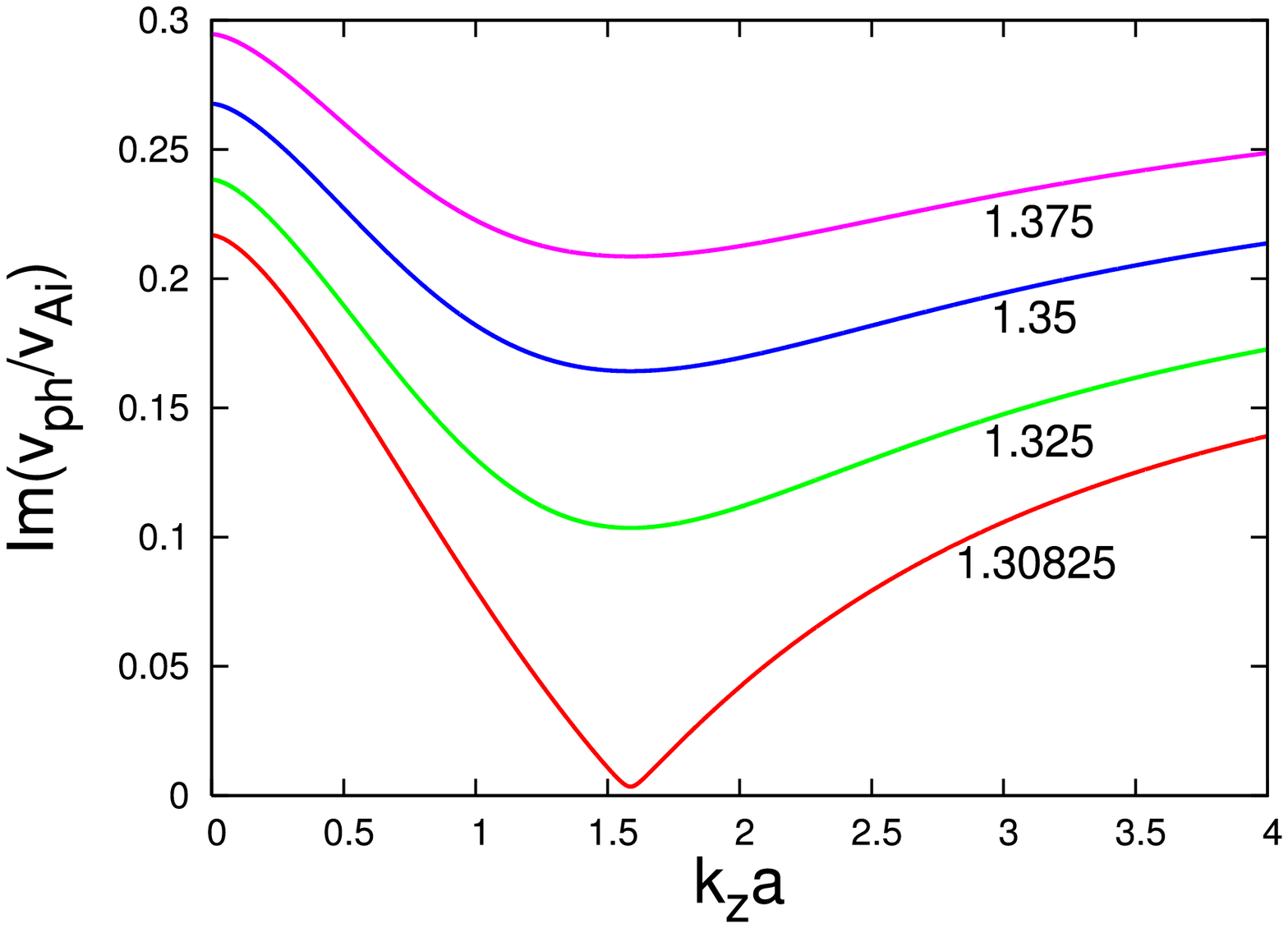}
  \end{tabular}
  \caption{(\emph{Left panel}) Dispersion curves of forward propagating stable
  and unstable kink waves in an untwisted magnetic flux tube at $\eta = 2$
  and $B_{\rm e}/B_{\rm i} = 0$ for
  various values of the Alfv\'en-Mach number $M_{\rm A}$. (\emph{Right panel})
  Growth rates of the unstable kink waves for $M_{\rm A} = 1.30825, 1.325, 1.35$,
  and $1.375$.}
  \label{fig:fig3}
\end{figure*}

Bennett et al.\ (\cite{bennett99}), in studying the kink wave propagation in
a twisted magnetic tube with parameters similar to ours, had discovered a
band that contains an infinite number of
pseudosurface (body) wave harmonics (Fig.~5(a) in their paper).
Among those, generally super-Alfv\'enic, waves we were able to easily
calculate only $11$ dispersion curves from Eq.~(\ref{eq:dispeq}) with search 
steps (in decreasing order) from $0.00025$ to $0.000000025$.  They
are not plotted in Fig.~\ref{fig:fig1} because there is no room for all of
them in that narrow band, but they will be shown shortly in a similar
plot for a tube with $\varepsilon = 0.4$.  With diminishing search step,
taking it equal to $0.0000000025$ or smaller, one can extract a few
($3$) additional curves, but finding them is an extremely hard
numerical task -- for such small search steps the data are very noisy.
However, these high-harmonics pseudosurface kink waves are not too
important to us since we are primarily interested in the evolution of
the kink mode associated with the kink speed $c_{\rm k}$ when the plasma
starts to flow.  It is worth noticing that the
\textsf{c}$_{\mathsf{k}}$-labelled dispersion curve for small
dimensionless wave numbers, till $k_z a = 0.1228$, describes a pseudosurface
(body) wave, while behind that value the wave is a pure surface mode.
These types of waves are sometimes called \emph{hybrid waves\/} (Bennett
et al.\ \cite{bennett99}).

It is essential to emphasise that the dispersion equation (\ref{eq:dispeq})
also yields solutions for negative values of the wave phase velocity (i.e.,
for backward-propagating waves), which are a mirror image of the 
solutions  in the positive semi-space.
Including the flow, the whole pattern shifts en bloc upwards, changing of
course both the \textsf{c}$_{\mathsf{k}}$-labelled dispersion curve and the
wave harmonics dispersion curves.  For each $M_{\rm A}$ we have \emph{two\/}
$c_{\rm k}$-dispersion curves that initially, for small Alfv\'en-Mach numbers,
are independent, but at certain magnitudes of $M_{\rm A}$ begin to interact;
for example, they form semi-closed or closed dispersion curves by merging.  These
dispersion curves are an indication that we are in a region
in which the kink wave may become unstable (see Zhelyazkov \cite{ivan10}).

To study the stability/instability status of the $c_{\rm k}$-kink wave, we have to
assume that the wave frequency is complex, i.e.,
$\omega \to \omega + \mathrm{i}\gamma$, where $\gamma$ is the expected
instability growth rate.  Thus, the dispersion equation becomes complex
(complex wave phase velocity and real wave number) and deriving the solutions
to a transcendental complex equation is generally a hard task (see Acton
\cite{acton90}).  We numerically solve Eq.~(\ref{eq:dispeq}) using the
Muller method (Muller \cite{muller56}) to find the complex roots at fixed
input parameters $\eta = 2$, $b = 0$, and $\varepsilon = 0.1$, and varying
Alfv\'en-Mach numbers $M_{\rm A}$ from zero to generally reasonable values.
The evolution of the two kink waves with increasing flow velocity (or
equivalently $M_{\rm A}$) is shown in Fig.~\ref{fig:fig2}.  We note that while
for rest
plasma or small Alfv\'en-Mach numbers the lower kink wave (i.e., the one that is
a mirror image of the green curve in Fig.~\ref{fig:fig1}) is a backward one,
for $M_{\rm A} = 1$ it becomes a forward wave.  At the next increase of the
Alfv\'en-Mach number the two dispersion curves change their form and structure
-- initially being hybrid modes, for $M_{\rm A} > 1.23$ they are pure surface
waves.  The most interesting observation is that for $M_{\rm A} \geqslant
1.2173$ both curves begin to merge and for $M_{\rm A} \geqslant 1.2933$ they
form a closed dispersion curve.  The ever increasing $M_{\rm A}$ yields yet
smaller closed dispersion curves -- the smallest one depicted in the left panel
of Fig.~\ref{fig:fig2} corresponds to $M_{\rm A} = 1.328$.  All these
dispersion curves present a stable propagation of the kink waves.  However, for
$M_{\rm A} \geqslant 1.32205$ we obtain a new family of wave dispersion curves
that correspond to an unstable wave propagation.  We plot in Fig.~\ref{fig:fig2}
four curves of that kind that were calculated for $M_{\rm A} = 1.32205$,
$1.347$, $1.372$, and $1.397$, respectively.  The growth rates of the unstable
waves are shown in the right panel of the same figure.  The instability that
arises is of the Kelvin-Helmholtz type.  Note the shape of red curve
labelled $\mathsf{1.32205}$: the growth rate has relatively high values for small
normalised wave numbers, then passes through a minimum (its relative value there is
$\sim\!\!0.003$) after which it starts to grow.  That specific curve is to some
extent a marginal curve -- for values of $M_{\rm A}$ lower than $1.32205$ the kink
wave is stable whilst at $M_{\rm A} \geqslant 1.32205$ the wave is definitely
unstable.  This means that the instability onset at the critical $M_{\rm A}$ will prevent
the occurrence of the neutral dispersion curve with $M_{\rm A} = 1.328$.

It is interesting to see how the twist of the magnetic field has changed the
dispersion curves and growth rates of unstable kink waves calculated for an
untwisted magnetic tube.  In that case the dispersion curves and growth
rates can be obtained through the exact numerical solutions of
Eq.~(\ref{eq:edwineq}) (see Zhelyazkov (\cite{ivan10})).  The results
are shown in Fig.~\ref{fig:fig3}.  Comparing Figs.~\ref{fig:fig2} and
\ref{fig:fig3}, we immediately see that the dispersion diagrams and growth
rates of the unstable kink waves are very similar.  There are differences, of
course; firstly the magnetic field twist visibly extends the closed dispersion curves in the horizontal direction, and secondly, it slightly increases the threshold
Alfv\'en-Mach number for the onset of the Kelvin-Helmholtz instability -- for
the untwisted magnetic tube it is equal to $1.30825$, whilst for the twisted one it
is slightly higher -- the corresponding value is $1.32205$.  The similarities
between the dispersion diagrams and the growth rates are important to us -- they
indirectly confirm the correctness of our numerical solutions derived by
solving the complex dispersion equation (\ref{eq:dispeq}).
\begin{figure}[ht]
   \centering
   \includegraphics[height=.23\textheight]{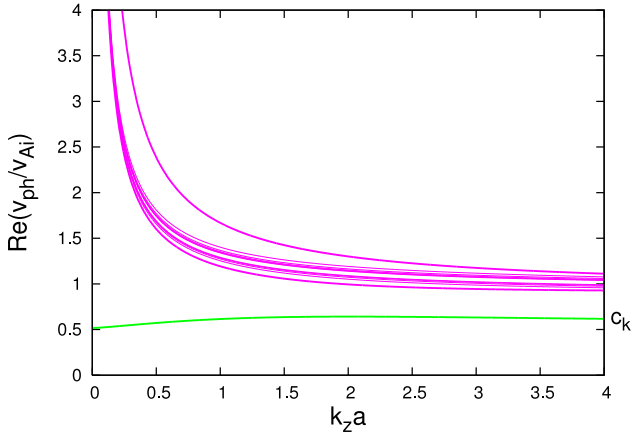}
   \caption{Dispersion curves of kink waves propagating along a twisted magnetic
   flux tube at $\eta = 2$, $B_{\rm e}/B_{\rm i} = 0$,
   $B_{{\rm i}\phi}/B_{\rm i} = 0.4$, and $M_{\rm A} = 0$.
   The green curve is the dispersion curve of the wave associated with the kink
   speed $c_{\rm k}$.  The purple curves depict all $11$ dispersion curves;
   $9$ of them are divided into two sub-groups.  The `evolution' of those curves
   when increasing $k_z a$ can be seen in the next figure, which shows them
   near the left top and right bottom corners of the plot.}
   \label{fig:fig4}
\end{figure}
\begin{figure}[ht]
\centering
  \begin{minipage}{\columnwidth}
  \centering
    \includegraphics[width=7.0cm]{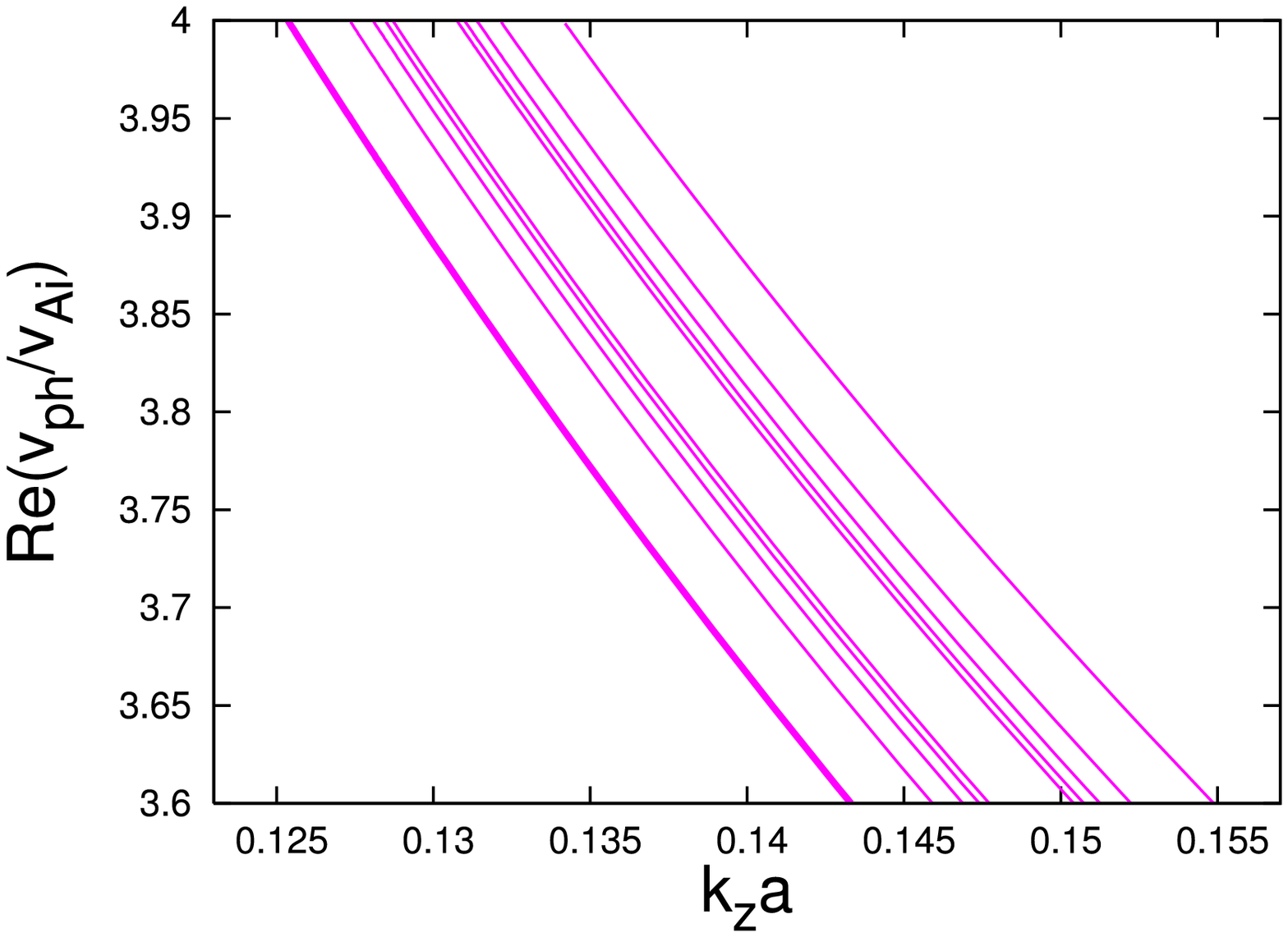} \\
\vspace{5mm}
    \includegraphics[width=7.0cm]{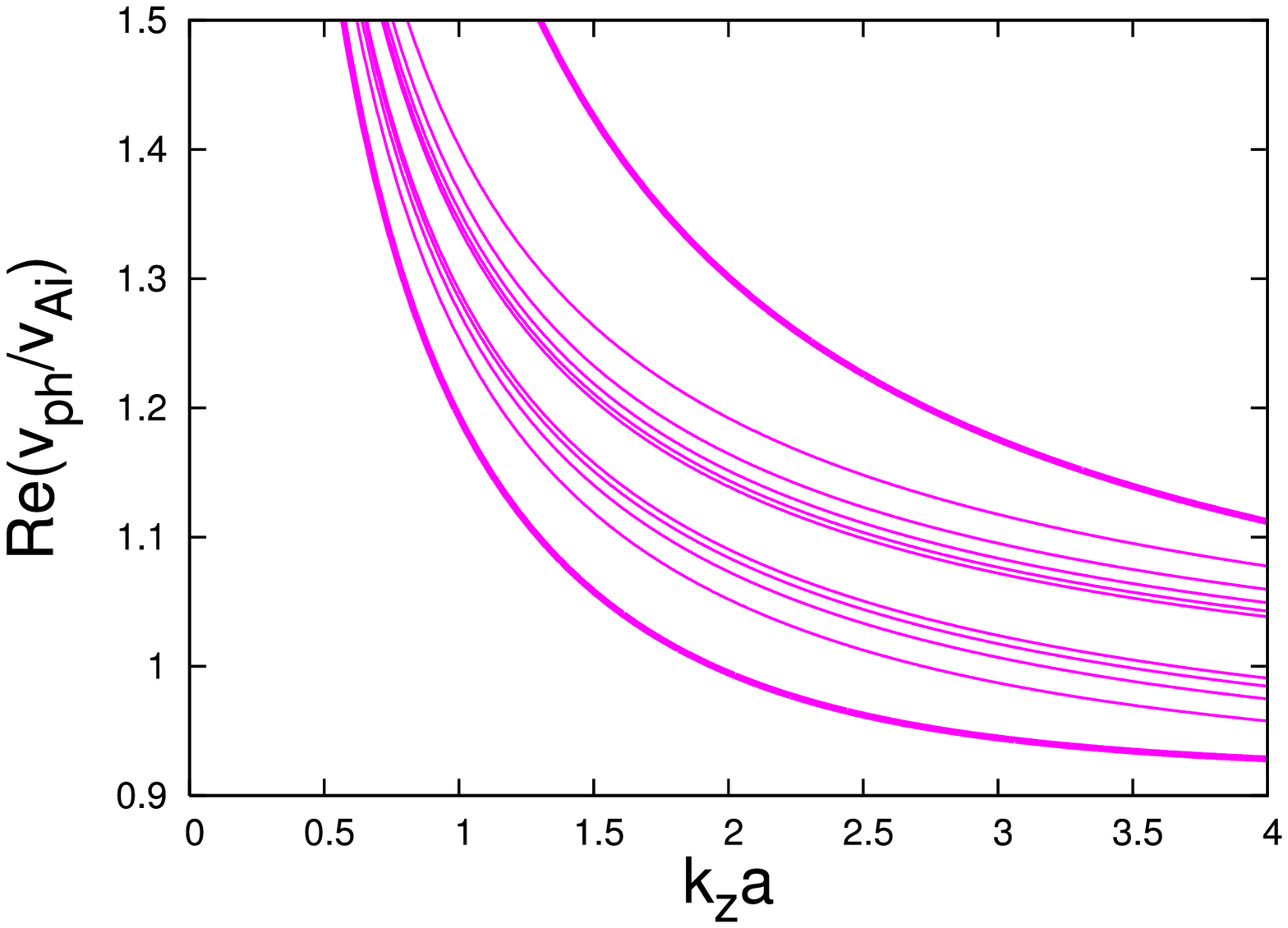}
  \end{minipage}
   \caption{Zoomed-in parts of the $11$ dispersion curves near the left top
   (\emph{top panel}) and right bottom (\emph{bottom panel}) corners of the
   plot in Fig~\ref{fig:fig4}.}
   \label{fig:fig5}
\end{figure}

Figure~\ref{fig:fig4} shows the dispersion curves of kink waves when the twist 
parameter $\varepsilon = 0.4$ and there is no flow.  Now one can see all
$11$ easily derived dispersion curves of the wave harmonics.  They have similar
shapes, and as seen in Figs.~\ref{fig:fig4} and \ref{fig:fig5}, one can
distinguish two sub-groups of four and five curves, respectively.  We note
that the eleventh (the highest) dispersion curve is not visible in the left
panel of Fig.~\ref{fig:fig5} because it simply begins at about
$k_z a = 0.23$, which is far beyond the right side of the plot.  Another
important observation is that the kink speed is markedly lower
than expected from the Eq.~(\ref{eq:kinkspeed}) value -- here its
normalised magnitude is equal to $0.5167$, which corresponds to a kink speed
of $5$ kilometers per second.  The kink wave itself is a hybrid mode -- till
$k_z a = 0.5623$ it is a pseudosurface (body) wave, becoming a pure surface
wave for the next values of the normalised wave number.  When plasma in the
tube is flowing, the kink waves are pure surface mode for $M_{\rm A} > 1.197$.

\begin{figure*}[ht]
  \centering
  \begin{tabular}{ccc}
    \includegraphics[width=6.5cm]{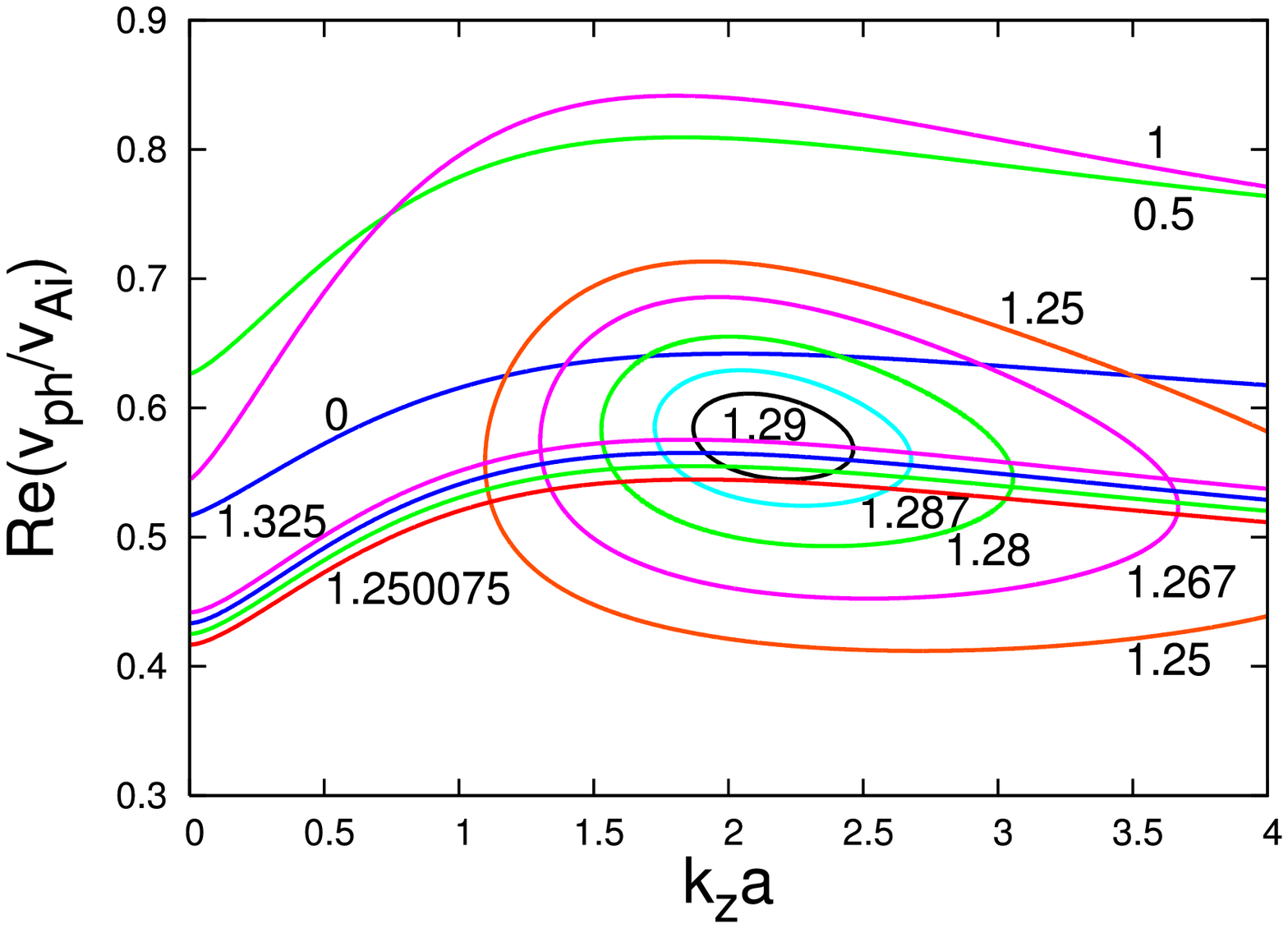} &
    \hspace{10mm} &
    \includegraphics[width=6.5cm]{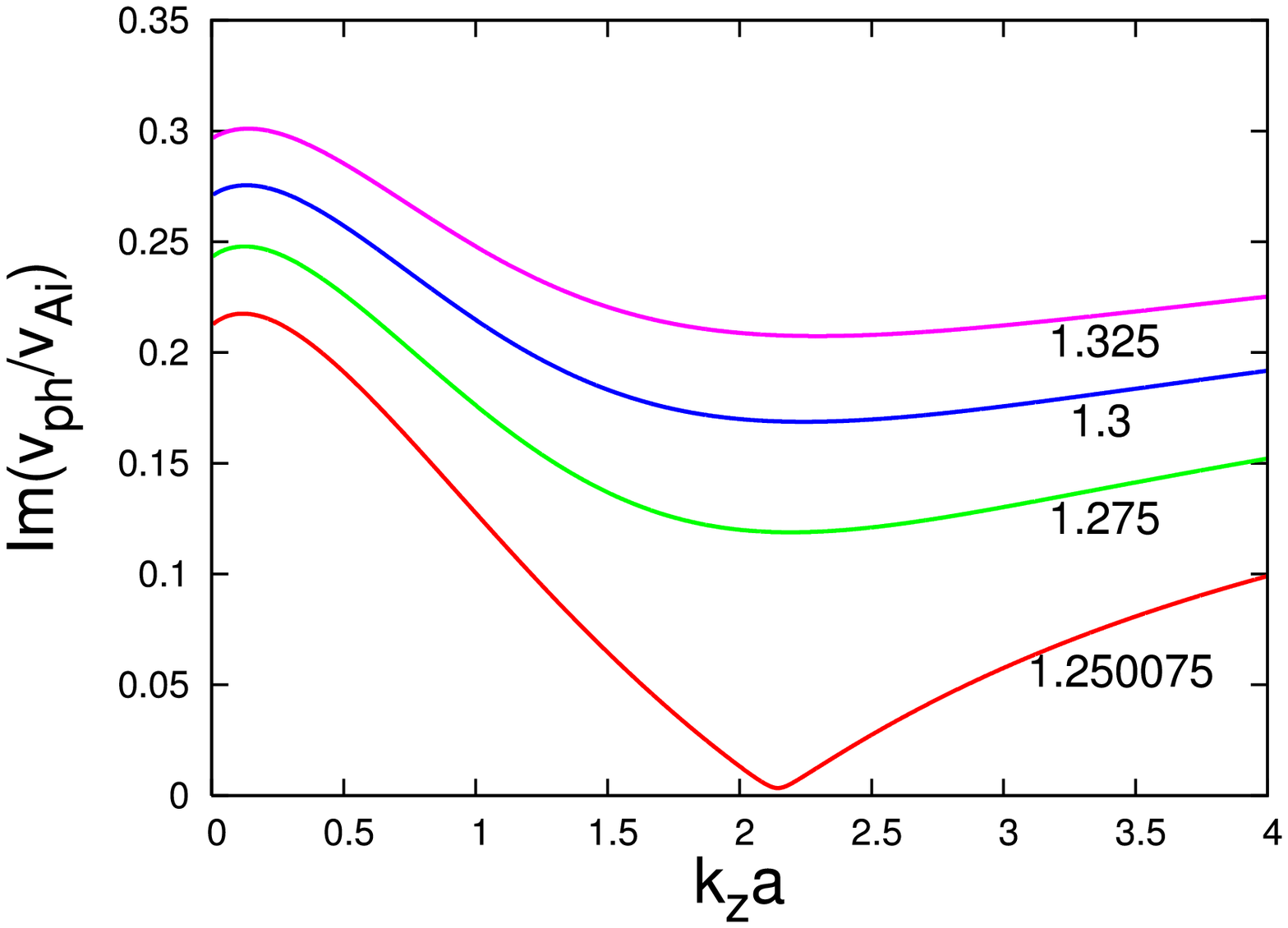}
  \end{tabular}
  \caption{(\emph{Left panel}) Dispersion curves of forward-propagating stable
  and unstable kink waves in a twisted magnetic flux tube with the same input
  parameters as in Fig.~\ref{fig:fig4} for various values of the Alfv\'en-Mach
  number $M_{\rm A}$.  (\emph{Right panel}) Growth rates of the unstable kink
  waves for $M_{\rm A} = 1.250075, 1.275, 1.3$, and $1.325$.}
   \label{fig:fig6}
\end{figure*}

The dispersion curves and growth rates of kink waves for this higher value of
the magnetic field twist are given in Fig.~\ref{fig:fig6}.  At the
left panel of this
figure we see dramatic changes in the shape and position of the closed
dispersion curves.  The most striking changes, however, are the circumstances that the
onset of unstable kink waves, like in Fig.~\ref{fig:fig2}, starts at
a value of the Alfv\'en-Mach number lower than that of the narrowest closed
dispersion curve -- $1.250075$ vs $1.29$!  In untwisted flux tubes the kink
mode appears at Alfv\'en-Mach numbers higher than those of the smallest closed
dispersion curves, as seen in Fig.~\ref{fig:fig3} (cf.\ also Fig.~4
in Zhelyazkov (\cite{ivan12a})).  In reality, of course, the onset of a
Kelvin-Helmholtz instability in a twisted tube will `reject' the
co-existence of stable kink waves like those labelled $\mathsf{1.267}$,
$\mathsf{1.28}$, $\mathsf{1.287}$, and $\mathsf{1.29}$ in Fig.~\ref{fig:fig6}.

\begin{figure*}[ht]
\centering
\begin{tabular}{ccc}
   \includegraphics[width=.31\textwidth]{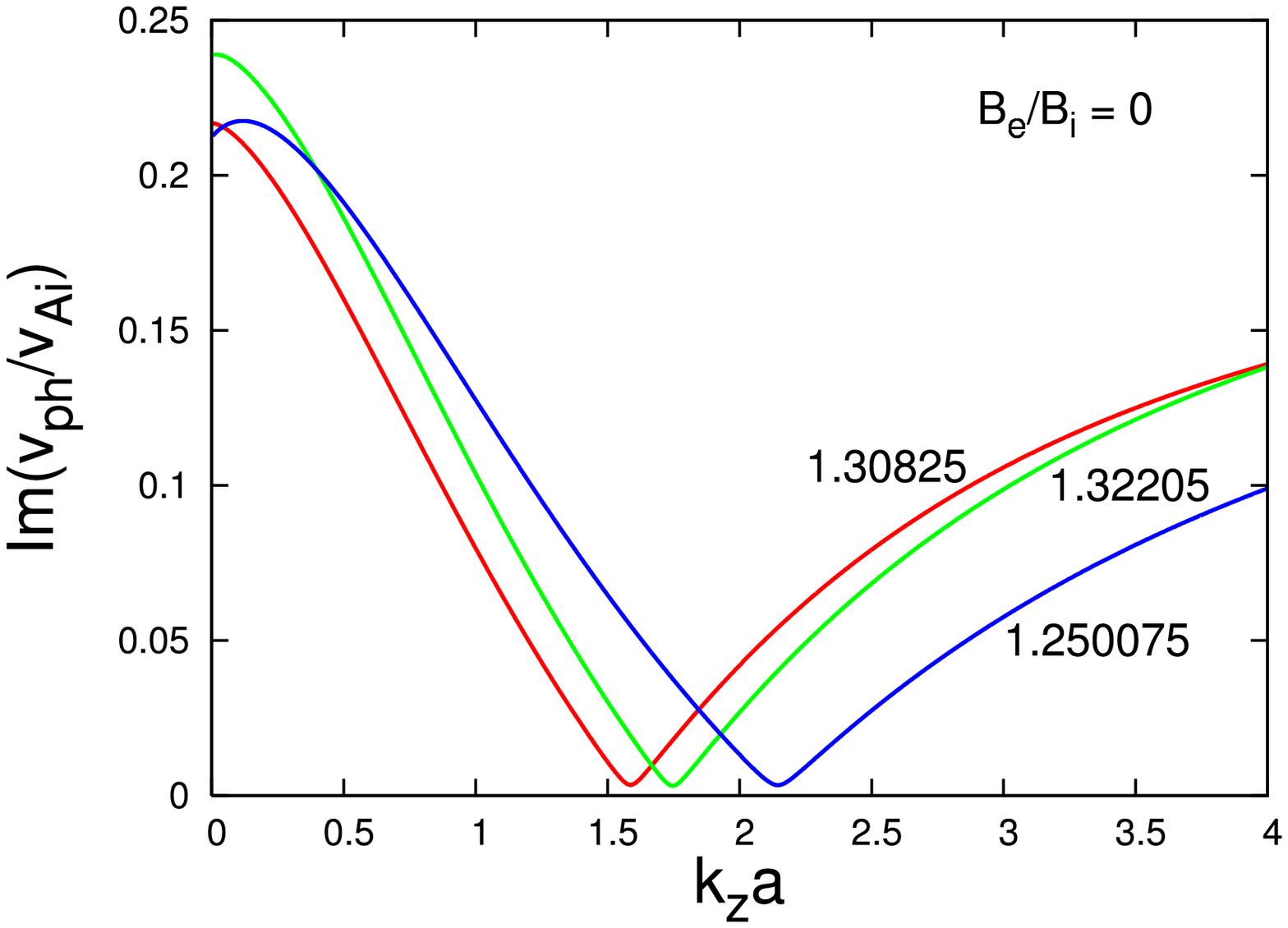} &
   \includegraphics[width=.31\textwidth]{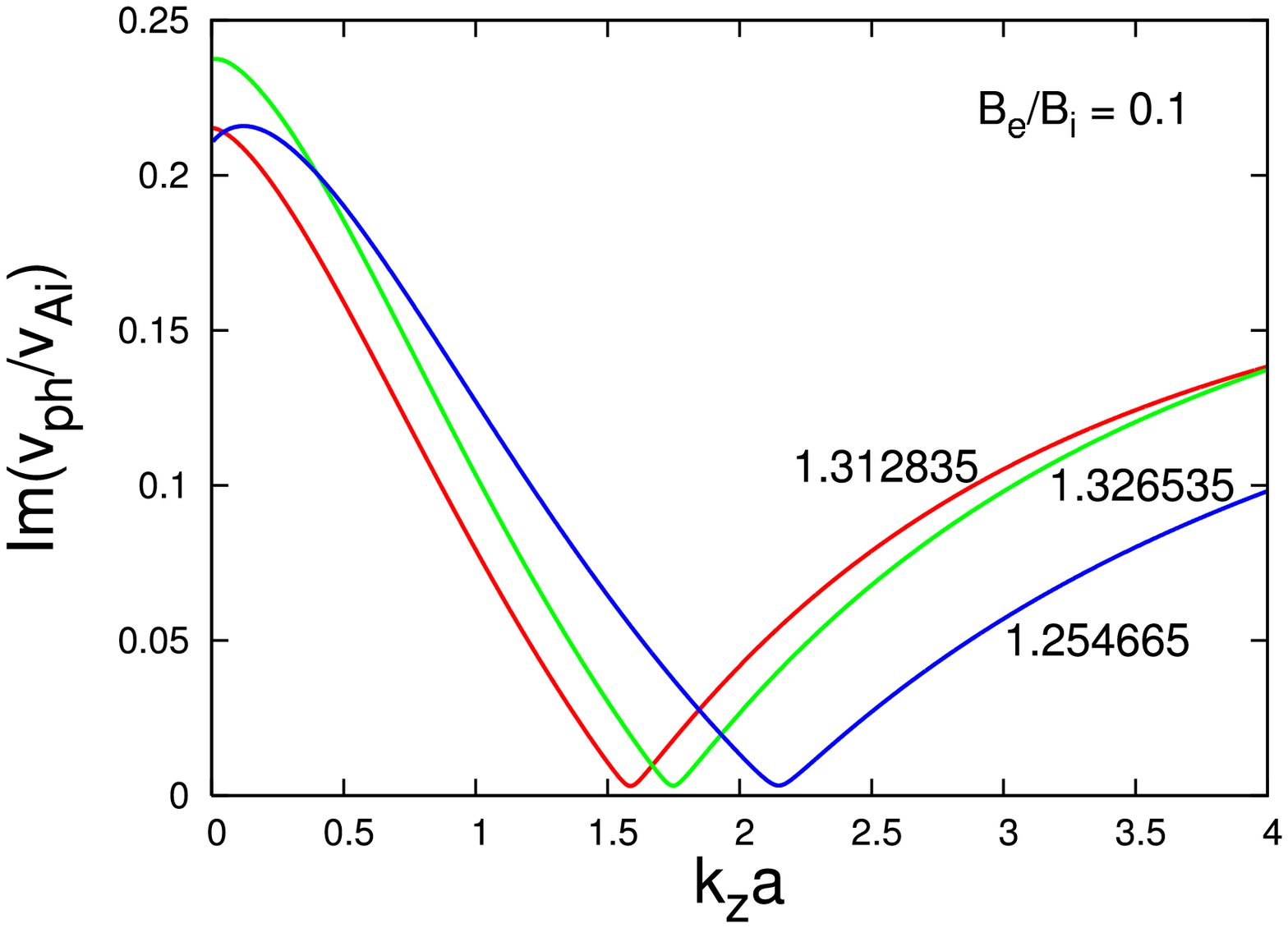} &
   \includegraphics[width=.31\textwidth]{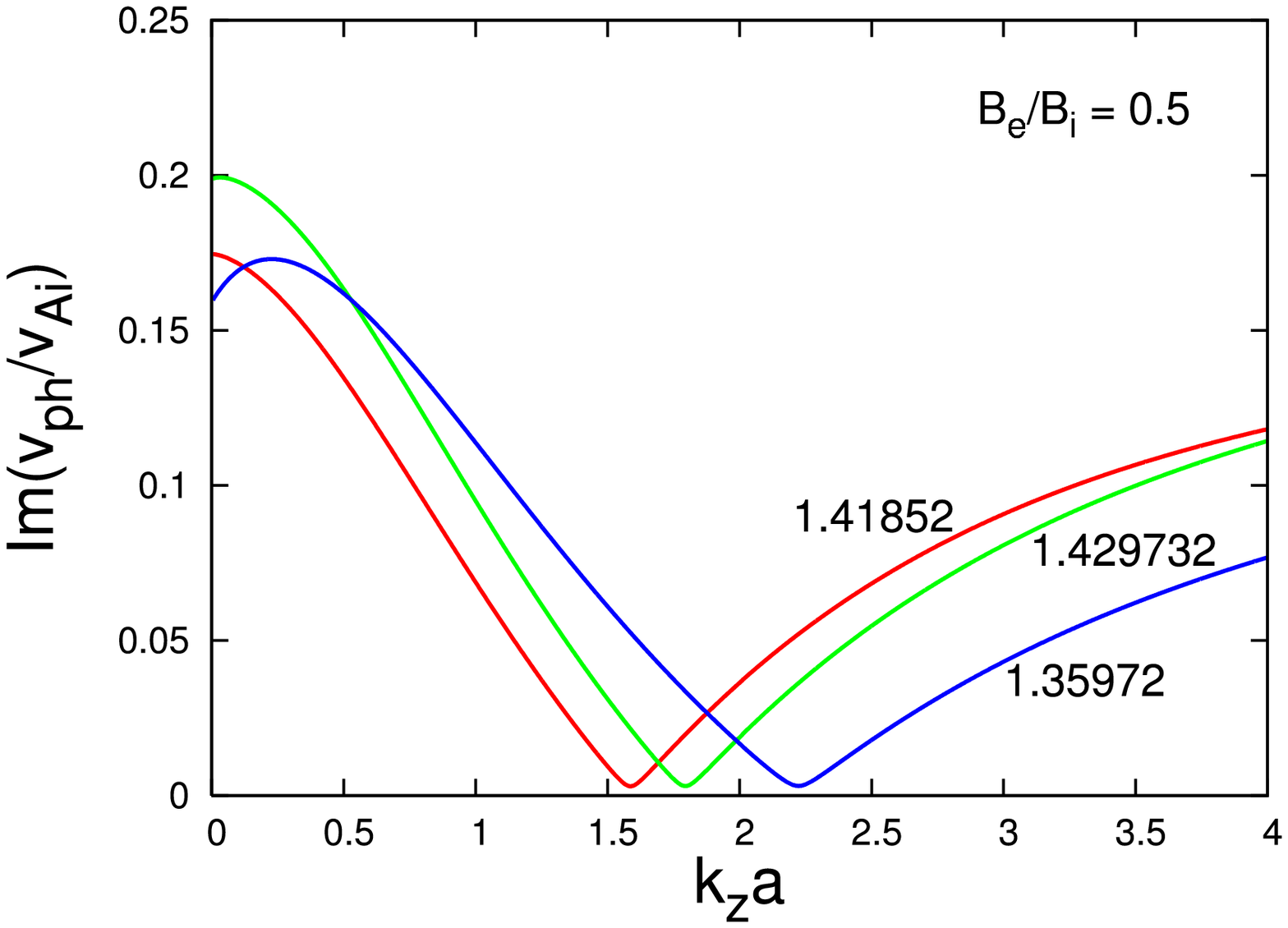}
\end{tabular}
   \caption{Specific growth rates of kink waves propagating along an
   untwisted/twisted magnetic flux tube at $\eta = 2$,
   $B_{\rm e}/B_{\rm i} = 0$ (\emph{left panel}),
   $B_{\rm e}/B_{\rm i} = 0.1$ (\emph{middle panel}), and
   $B_{\rm e}/B_{\rm i} = 0.5$ (\emph{right panel}) for various values
   of the twist parameter $B_{{\rm i}\phi}/B_{\rm i}$, equal to $0$ 
   (red curve), $0.1$ (green curve), and $0.4$ (blue curve), respectively.  The
   curve labels denote the threshold Alfv\'en-Mach number.}
   \label{fig:fig7}
\end{figure*}

\begin{figure*}[ht]
\centering
\begin{tabular}{ccc}
   \includegraphics[width=.31\textwidth]{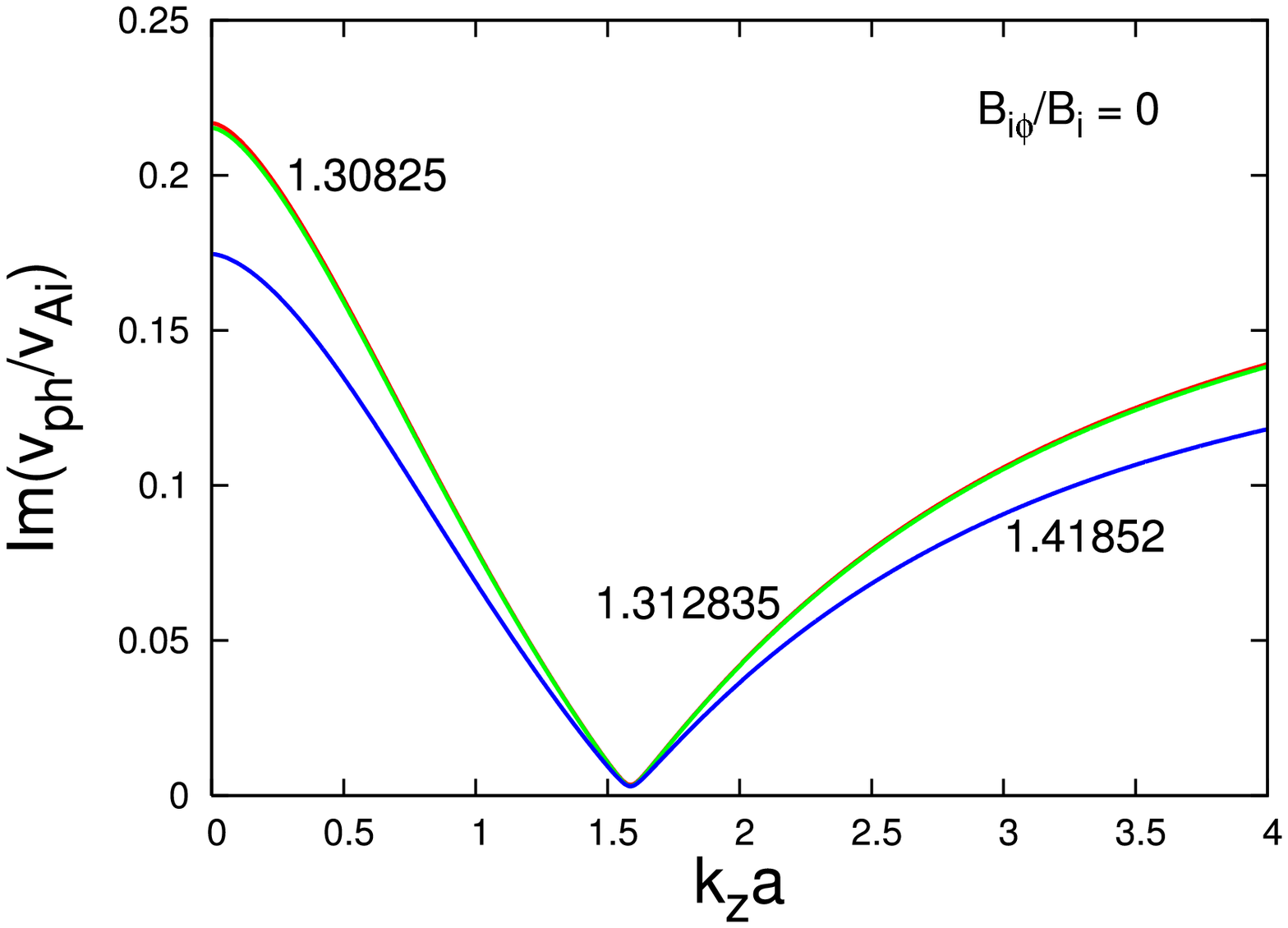} &
   \includegraphics[width=.31\textwidth]{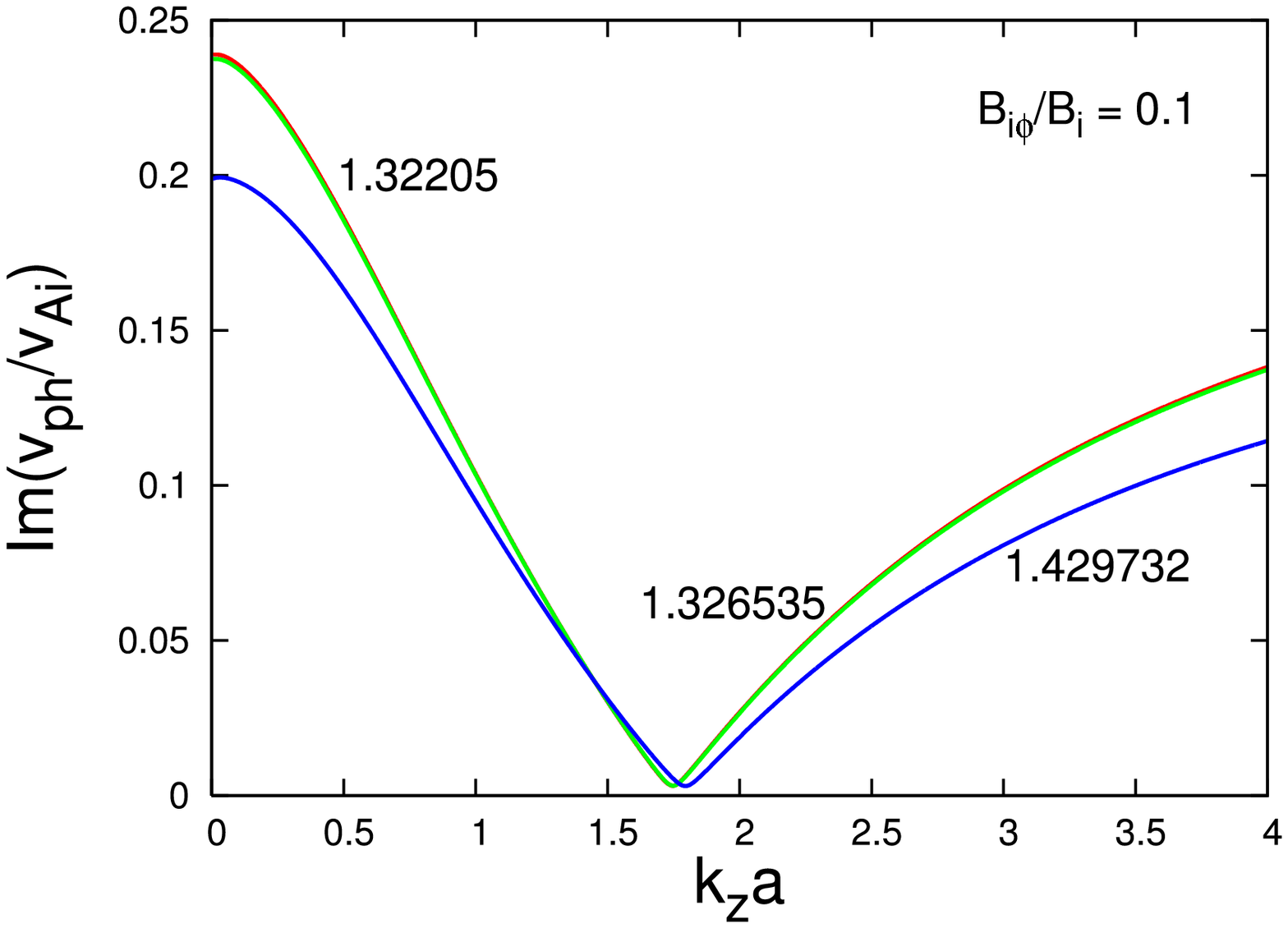} &
   \includegraphics[width=.31\textwidth]{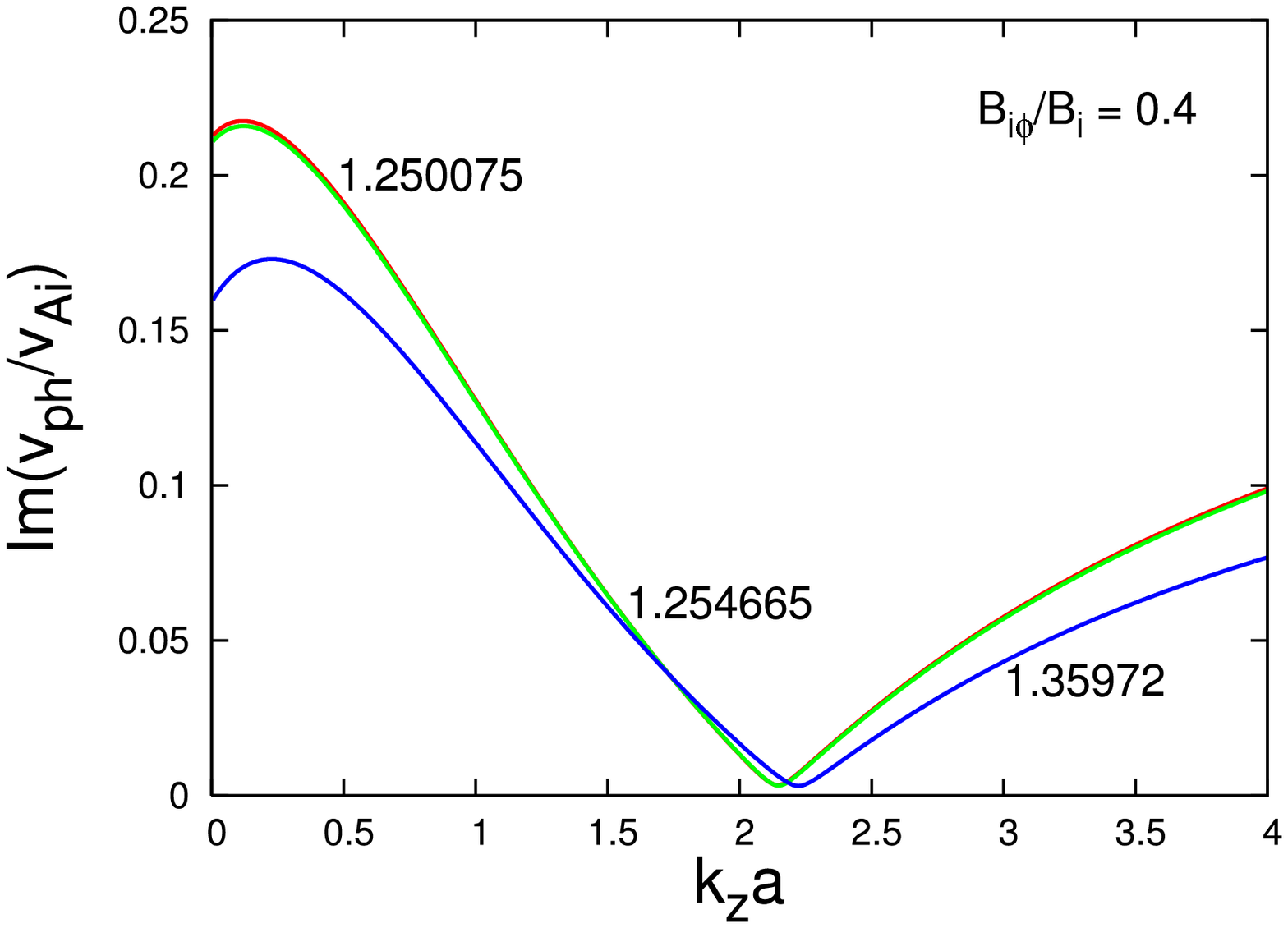}
\end{tabular}
   \caption{Specific growth rates of kink waves propagating along an
   untwisted/twisted magnetic flux tube at $\eta = 2$,
   $B_{{\rm i}\phi}/B_{\rm i} = 0$ (\emph{left panel}),
   $B_{{\rm i}\phi}/B_{\rm i} = 0.1$ (\emph{middle panel}), and
   $B_{{\rm i}\phi}/B_{\rm i} = 0.4$ (\emph{right panel}) for various values
   of the environment's magnetic field $B_{\rm e}/B_{\rm i}$, equal to $0$
   (red curve), $0.1$ (green curve), and $0.5$ (blue curve), respectively.  The
   curve labels denote the threshold Alfv\'en-Mach number.}
   \label{fig:fig8}
\end{figure*}

Similar calculations were performed for the two external magnetic
fields specified by $b = 0.1$ and $0.5$.  As expected, the
environment's magnetic field stabilises the kink wave propagation
(Bennett et al.\ \cite{bennett99}) -- the
instability onset occurs at slightly higher threshold Alvf\'en-Mach
numbers, but the shapes of the dispersion and growth rate curves are
similar to those of an isolated twisted flux tube as depicted in
Figs.~\ref{fig:fig2}, \ref{fig:fig3}, and \ref{fig:fig6}.  The influence
of the magnetic field twist, $\varepsilon$, at a fixed external magnetic
field, $b$, and that of the environment's magnetic field at a fixed tube
twist on the specific growth rate curves are illustrated in
Figs.~\ref{fig:fig7} and \ref{fig:fig8}.
As seen in Fig.~\ref{fig:fig7}, at a fixed magnetic field
configuration, the tube twist shifts to the right the minimum of the specific
growth rate curve.  This is most pronounced at $b = 0.5$.  A twist of $0.4$
notably decreases the wave growth rate for each $b$ and yields
the lowest threshold Alfv\'en-Mach number for a given $b$.  Thus, for a
photospheric twisted magnetic tube with $\varepsilon = 0.4$ flow speeds in the
range $12.5$--$13.6$~km\,s$^{-1}$ can ensure the occurrence of a Kelvin-Helmholtz
instability of the kink waves propagating along that tube.  These flow speeds
can be observed/detected in the solar photosphere.  It becomes clear from
Fig.~\ref{fig:fig8} that an external magnetic field visibly increases the
threshold Alfv\'en-Mach number and diminishes the wave growth rate -- the latter
effect is strongest for a magnetic tube with twist parameter $\varepsilon = 0.4$.

\subsection{Kink waves in twisted magnetic tubes with a density contrast
            $\eta = 0.1$}
\label{subsec:eta01}

\begin{figure*}[ht]
  \centering
  \begin{tabular}{ccc}
    \includegraphics[width=6.5cm]{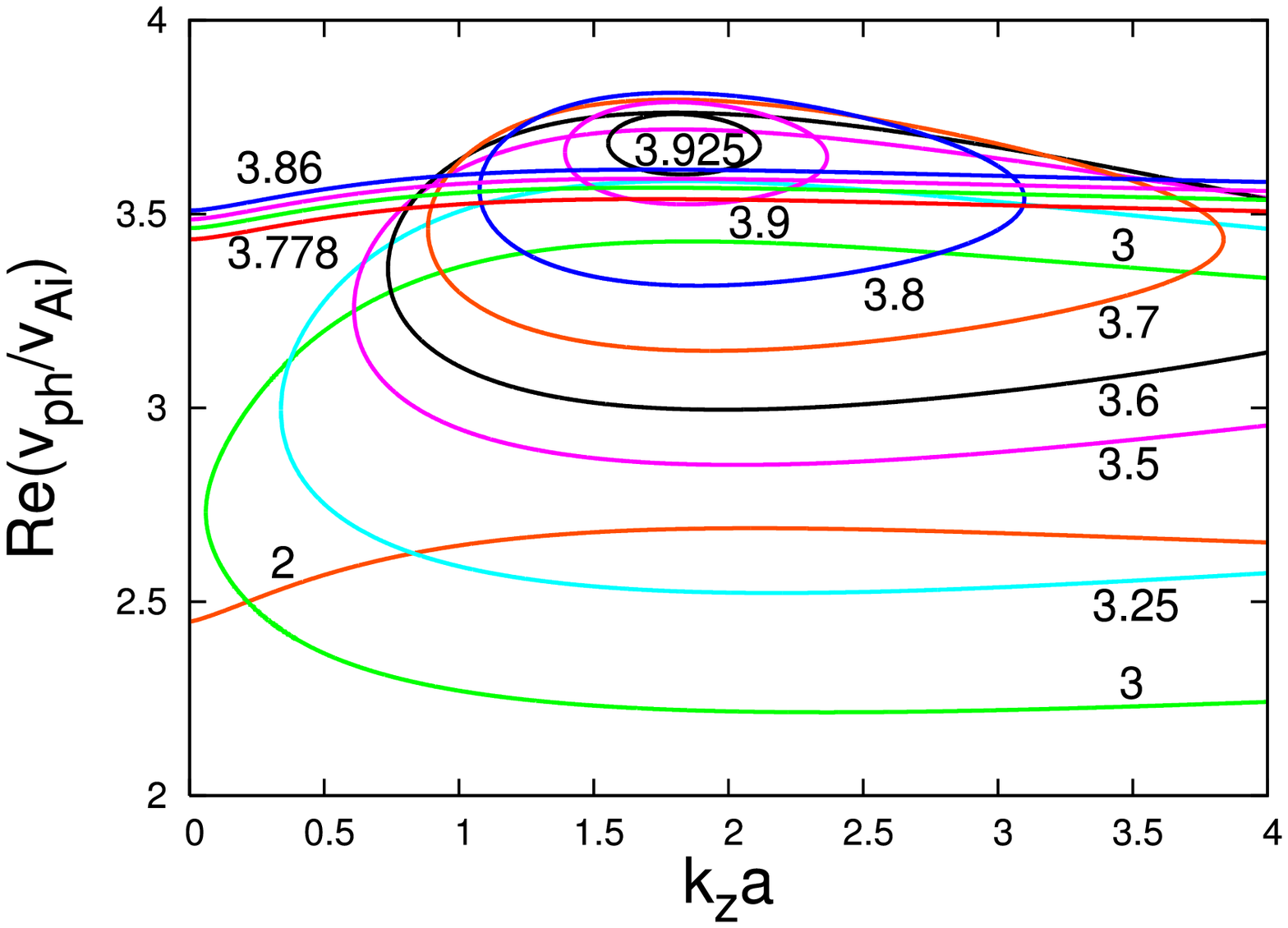} &
    \hspace{10mm} &
    \includegraphics[width=6.5cm]{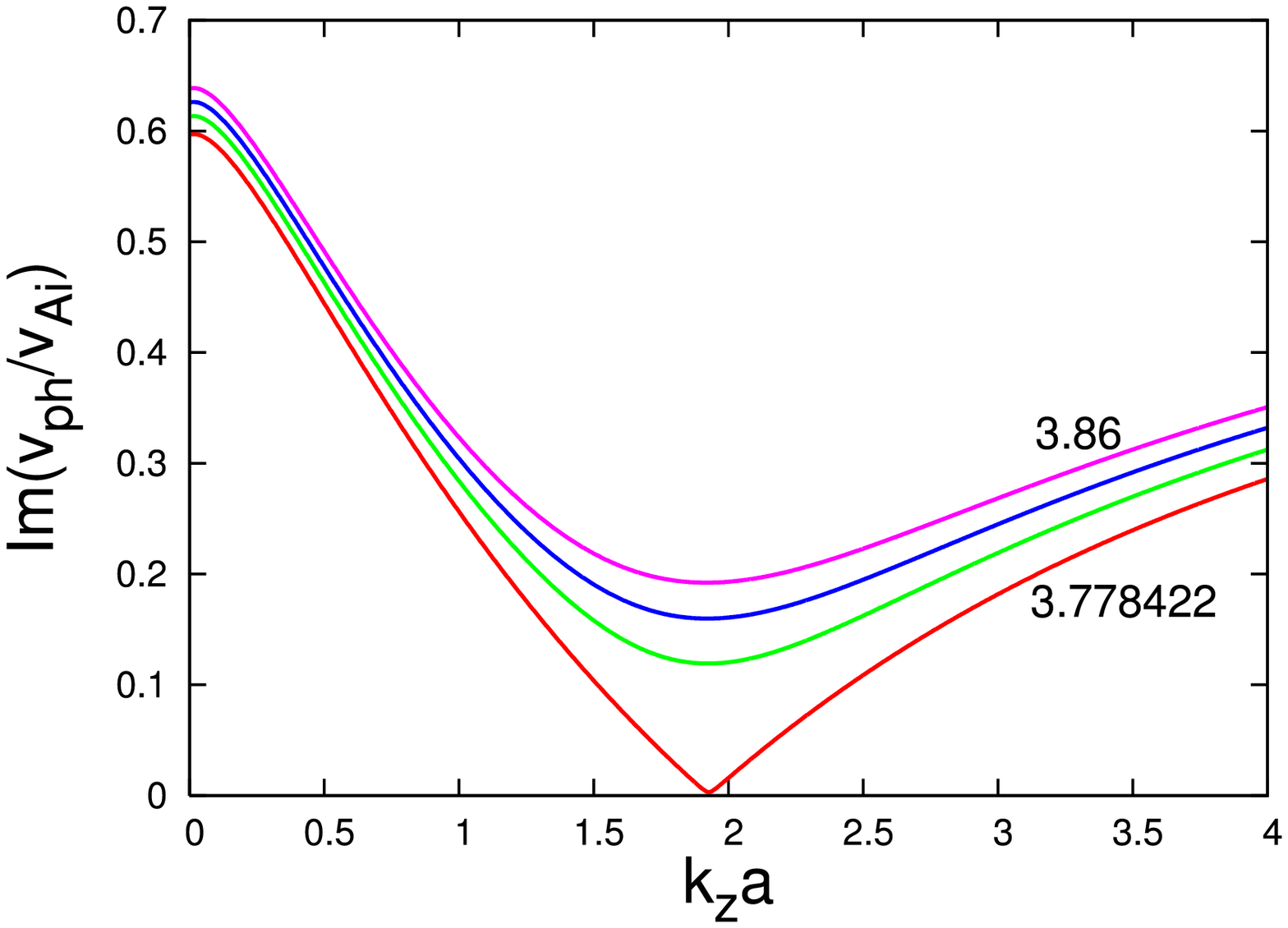}
  \end{tabular}
  \caption{(\emph{Left panel}) Dispersion curves of forward-propagating stable
  and unstable kink waves in a twisted magnetic flux tube with $\eta = 0.1$,
  $B_{\rm e}/B_{\rm i} = 0$, and $B_{{\rm i}\phi}/B_{\rm i} = 0.4$ for various
  values of the Alfv\'en-Mach number $M_{\rm A}$.  (\emph{Right panel}) Growth
  rates of the unstable kink waves for $M_{\rm A} = 3.778422, 3.81, 3.835$,
  and $3.86$.}
  \label{fig:fig9}
\end{figure*}

\begin{figure*}[ht]
\centering
\begin{tabular}{ccc}
   \includegraphics[width=.31\textwidth]{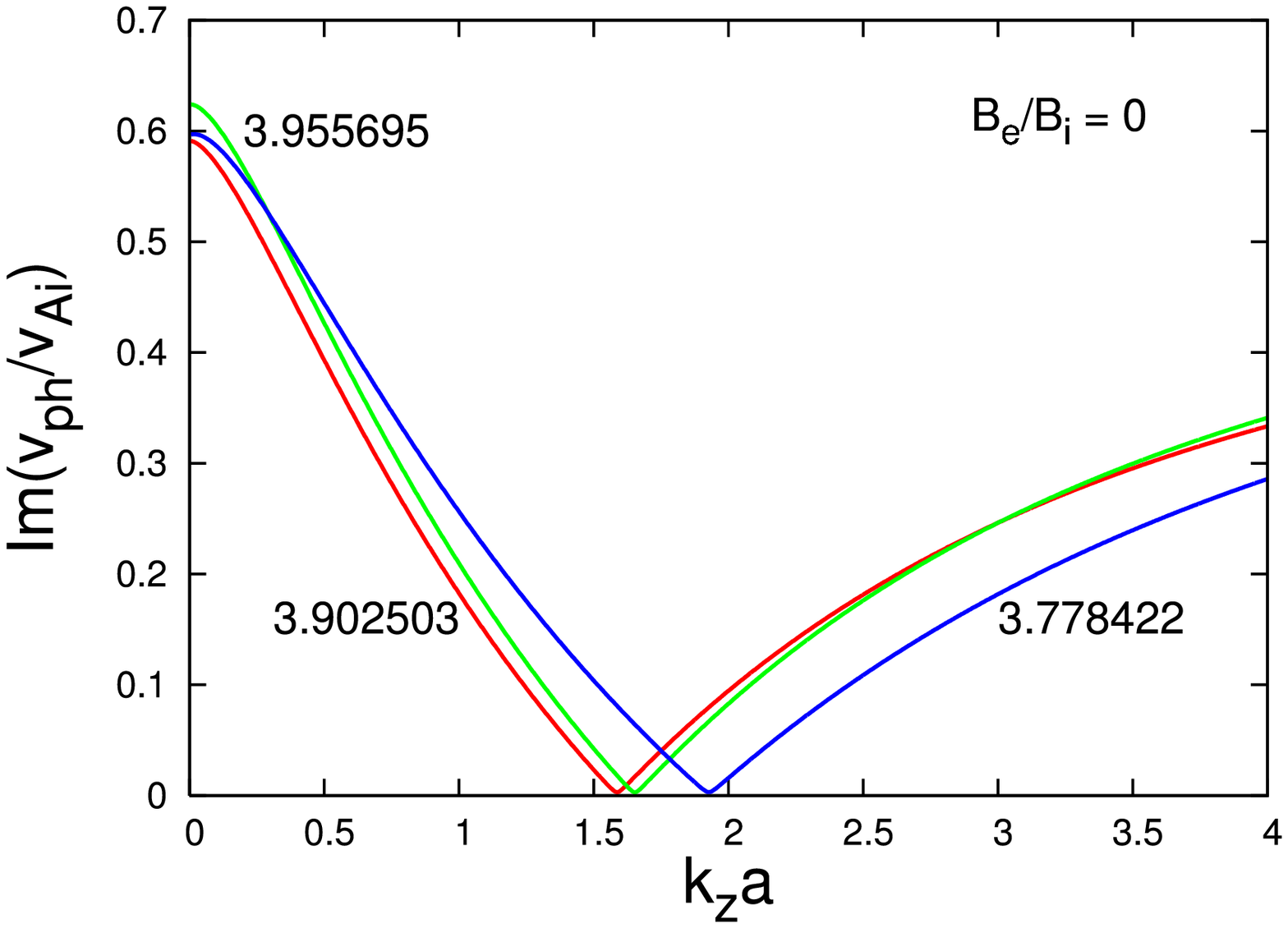} &
   \includegraphics[width=.31\textwidth]{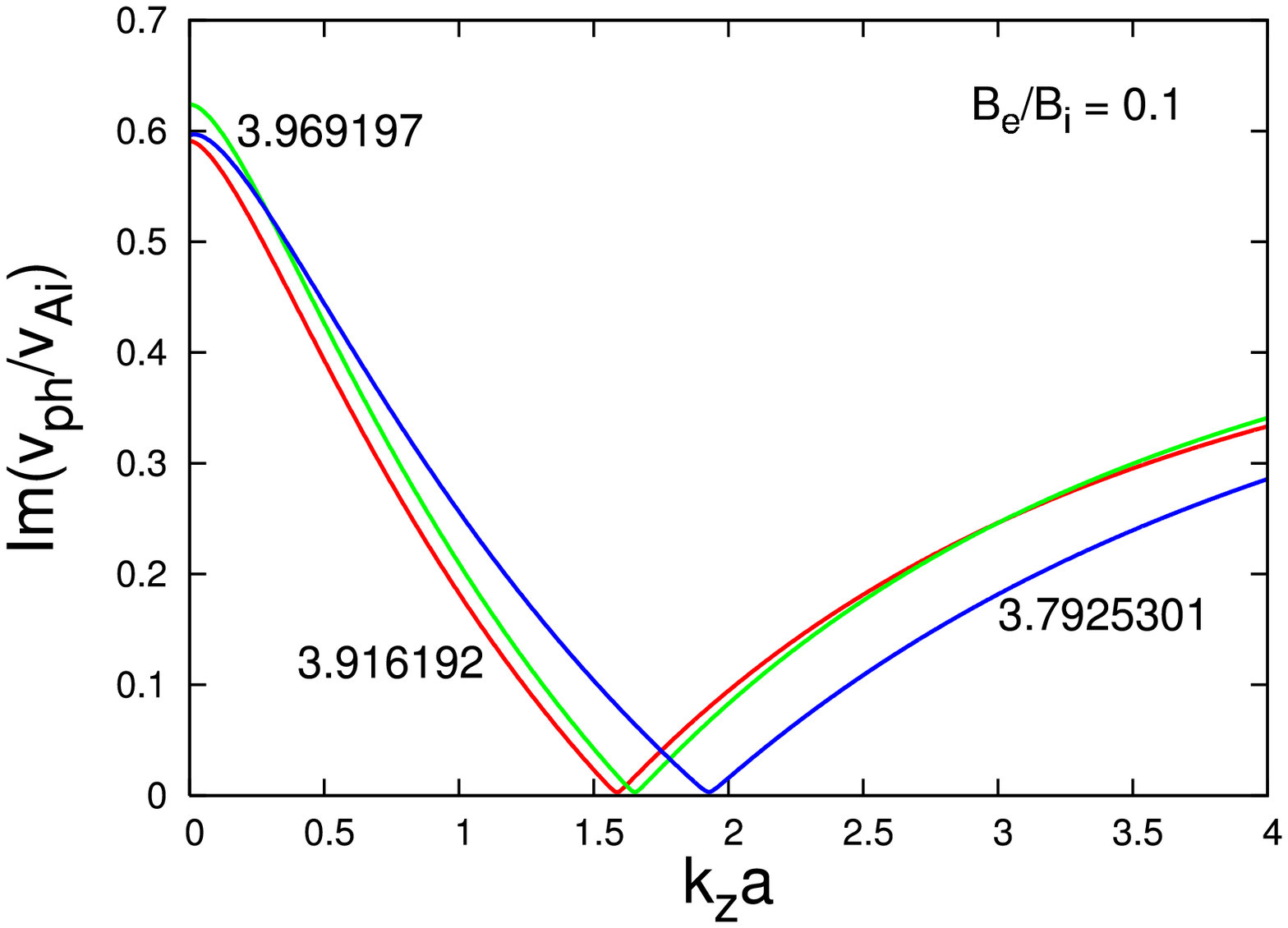} &
   \includegraphics[width=.31\textwidth]{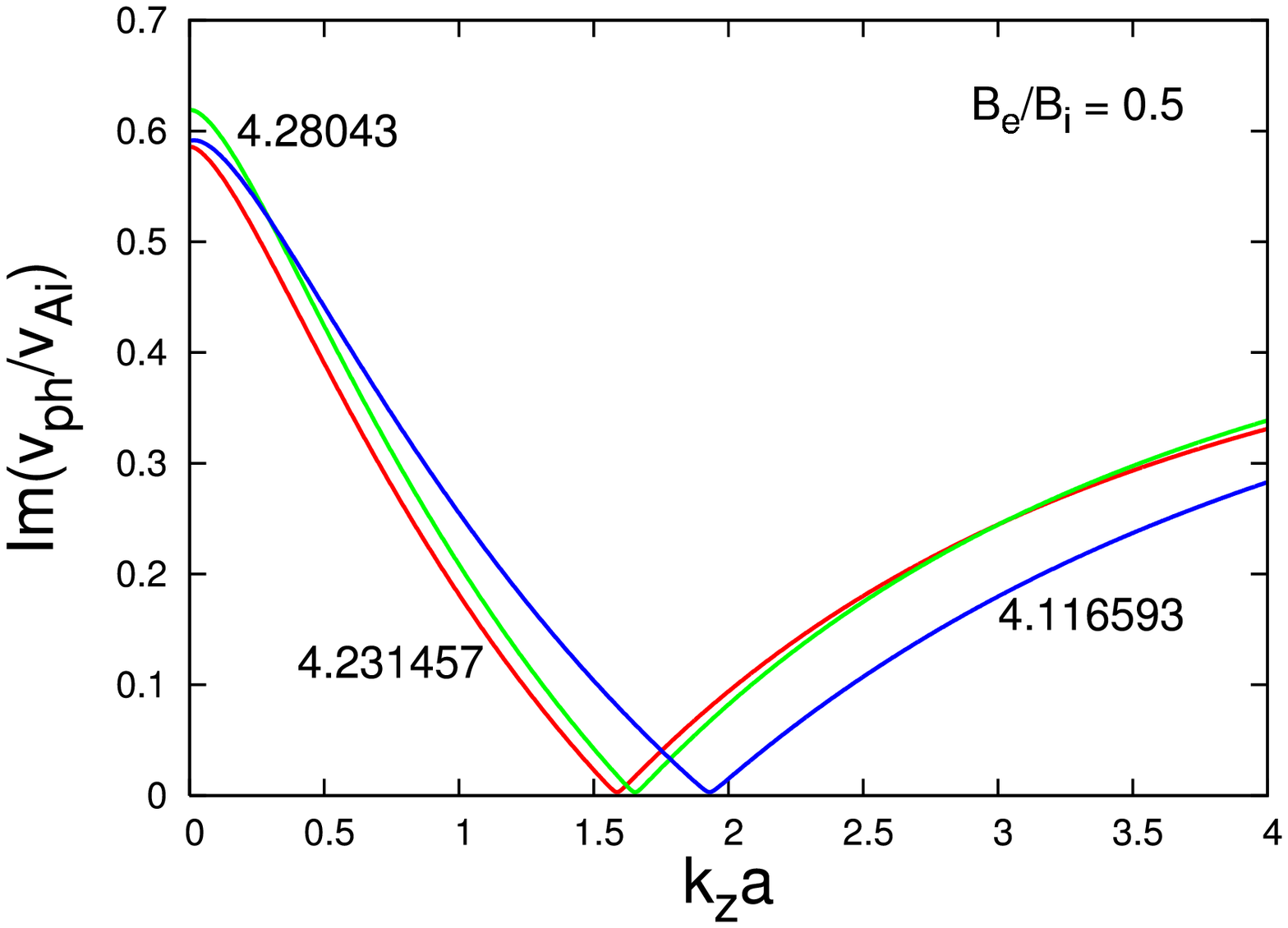}
\end{tabular}
   \caption{Specific growth rates of kink waves propagating along an
   untwisted/twisted magnetic flux tube at $\eta = 0.2$,
   $B_{\rm e}/B_{\rm i} = 0$ (\emph{left panel}),
   $B_{\rm e}/B_{\rm i} = 0.1$ (\emph{middle panel}), and
   $B_{\rm e}/B_{\rm i} = 0.5$ (\emph{right panel}) for various values
   of the twist parameter $B_{{\rm i}\phi}/B_{\rm i}$, equal to $0$
   (red curve), $0.1$ (green curve), and $0.4$ (blue curve), respectively.  The
   curve labels denote the threshold Alfv\'en-Mach number.}
   \label{fig:fig10}
\end{figure*}

A density contrast $\eta = 0.1$ is perhaps more suitable for a coronal and
upper chromospheric magnetic flux tube than to a photospheric one, but
nevertheless, studying this case is important for us to see how the
density contrast affects the threshold/critical Alfv\'en-Mach numbers for the 
onset of the Kelvin-Helmholtz instability of kink waves propagating along twisted
magnetic tubes.  The procedure for performing calculations and
plotting dispersion diagrams is the same as in the previous subsection.
This lower value of $\eta$ yields markedly higher threshold
Alfv\'en-Mach numbers as well as higher growth rates compared to
$\eta = 2$.  Here, in Fig.~\ref{fig:fig9}, we show only one set of
plots -- namely that for an isolated flux tube ($b = 0$) with a twist of
$0.4$.  At these values of the input parameters we obtain the smallest
threshold $M_{\rm A}$.  In Fig.~\ref{fig:fig10}, like in Fig.~\ref{fig:fig7},
we present the specific growth rate curves at fixed external magnetic fields
for the three different values of the magnetic twist.  As seen from
this figure, for that value of the density contrast ($\eta = 0.1$) the
influence of the magnetic twist, $\varepsilon$, on the shapes and positions
of the specific growth rate curves, irrespective of the magnitude of the
environment's magnetic field, is negligibly small.  That is why a
figure, similar to Fig.~\ref{fig:fig8}, is useless -- in each subfigure for
given $\varepsilon$ the three curves would practically coincide because the
differences are only in the second/third places behind the decimal point.
As for a photospheric twisted tube, one obtains the lowest
threshold $M_{\rm A}$ at $b = 0$ and $\varepsilon = 0.4$ -- the
corresponding value is $3.778422$.  If we accept that such a twisted
magnetic tube can model a coronal loop, to estimate the Alfv\'en
speed inside the tube, we assume a typical particle number density
$n_{\rm i} \approx 10^{9}$ cm$^{-3}$ (see Ugarte-Urra et al.\ \cite{ugarte05})
and a magnetic field strength $B_{\rm i} = 11$ G (see Verwichte et al.\
\cite{erwin09}); then $v_{\rm Ai} \approx 900$ km\,s$^{-1}$.  With this
high Alfv\'en speed one would need a flow speed of about
$3400$~km\,s$^{-1}$ for the onset of an instability of the Kelvin-Helmholtz
type.  Since the detected flow velocities in the coronal loops are of the order
of a few tens of kilometers per second, we can conclude that the kink waves
in coronal loops with axial mass flows are stable against the Kelvin-Helmholtz
instability.

\section{Conclusion}
\label{sec:concl}

We have studied the
dispersion properties and the stability of the MHD kink modes running along the
length of photospheric/chromospheric uniformly twisted magnetic tubes with
axial mass flows.  These were modelled as straight cylindrical jets
of ideal incompressible plasma surrounded by a rest unmagnetised/magnetised also
incompressible fully ionized medium.  The wave propagation was
investigated in the context of standard ideal magnetohydrodynamics by using
linearised equations for the perturbations of the basic quantities: pressure,
fluid velocity, and wave magnetic field.  The derived dispersion equation
describes the propagation of the kink mode influenced by the twist of the
equilibrium magnetic field $\vec{B}_{\rm i}$ and the presence of moving
plasma.  The twist itself is characterised by the ratio of the azimuthal,
$B_{{\rm i}\phi}$, to the axial, $B_{{\rm i}z}$, background magnetic field
components both evaluated at the inner surface of the twisted magnetic tube.
To avoid competition from the so-called kink instability, which is largely
caused by the twist of the equilibrium magnetic field, we chose the ratio
$B_{{\rm i}\phi}/B_{{\rm i}z}$ to be less than $1$.  For simplicity and clarity
of the numerical code used for solving the wave dispersion relation (in
complex variables) we used a safer twist parameter $\varepsilon$ defined
as $B_{{\rm i}\phi}/B_{\rm i}$.  Our $\varepsilon$ and the real twist
$B_{{\rm i}\phi}/B_{{\rm i}z}$ have generally rather close magnitudes,
because they are related by a simple expression.  The streaming plasma is
characterised by its velocity $\vec{v}_0$, which is directed along the
axis of the twisted magnetic tube.  An alternative and more convenient way
of specifying the mass flow is by defining the Alfv\'en-Mach number: the
ratio of the jet speed to the Alfv\'en speed inside the jet,
$M_{\rm A} = v_0/v_{\rm Ai}$.  The key parameters controlling the dispersion
properties of the kink waves are the so-called density contrast,
$\eta = \rho_{\rm e}/\rho_{\rm i}$, defined as the ratio of the density of
the environment to that of the twisted tube, the ratio of the two
background magnetic fields, $b = B_{\rm e}/B_{\rm i}$, the twist of the
magnetic field, $\varepsilon$, and, naturally, the mass flow characterised
by the Alfv\'en-Mach number $M_{\rm A}$.

To study the influence of the mass flow, the twist of the equilibrium
magnetic field inside the tube, as well as the environment's magnetic
field on the dispersion properties and stability of the kink waves,
we considered two cases with fixed density contrasts $\eta = 2$ and
$0.1$, respectively.  For a given $\eta$ and a specified external
magnetic field via the parameter $b$, we investigated the kink wave
dispersion curves and the instability growth rate when the wave becomes
unstable for two different magnetic field twist parameters, namely
$\varepsilon = 0.1$ and $0.4$, respectively, by gradually changing the
normalised mass flow velocity, i.e, the Alfv\'en-Mach number $M_{\rm A}$,
from zero (rest plasma) to some reasonable values.  The flow shifts
both the kink-speed curves and the dispersion curves of the wave harmonics
upwards.  We focused our study on the kink-speed wave only since
it is the mode that plays an important role in the dynamics of photospheric
and chromospheric/coronal flows.

To evaluate the effect of the adopted magnetic field twist, $\varepsilon$, on
the kink-speed wave dispersion curves and chiefly on the value of the critical
Alfv\'en-Mach number that determines that flow speed that ensures the onset of
an instability of the Kelvin-Helmholtz type, we compared the wave dispersion
diagrams in twisted tubes with those in untwisted magnetic tubes.  It is well
established that in untwisted magnetic tubes the kink waves are subject to the
Kelvin-Helmholtz instability when $M_{\rm A}$ exceeds a certain critical value,
depending upon the density contrast, $\eta$, and on the ratio of the
background magnetic fields, $b$, in both media (see
Vasheghani Farahani et al.\ \cite{farahani09}; Zhelyazkov \cite{ivan10,ivan12a}).
For an isolated magnetic tube, $b = 0$, and, hence,
the threshold value of the Alfv\'en-Mach number depends only on the density
contrast, $\eta$.  Relatively high values of $\eta$ yield lower threshold
Alfv\'en-Mach numbers while the lower values of $\eta$ do just the
opposite -- for our chosen two $\eta$s the corresponding critical normalised
flow speeds are $1.30825$ and $3.902503$.  The effect of the magnetic field
twist critically depends on the value of the twist parameter $\varepsilon$.
This effect is strongest when $\varepsilon = 0.4$ -- then one observes
a noticeable decrease in the threshold Alfv\'en-Mach number: for a tube with
$\eta = 2$ and $b = 0$,
$M_{\rm A}^{\rm cr} = 1.250075$, while at $\eta = 0.1$ it is equal to $3.778422$.
This substantial finding that the twisted magnetic field lowers the
threshold $M_{\rm A}$ for the occurrence of the Kelvin-Helmholtz
instability of $m = 1$ mode agrees with the same statement declared
in the works by Bennett et al. (\cite{bennett99}) and Zaqarashvili et al.\
(\cite{temury10}).  In particular, in twisted photospheric magnetic
tubes with a density contrast $\eta = 2$, magnetic field twist $\varepsilon =
0.4$, and Alfv\'en speed $v_{\rm A} = 10$ km\,s$^{-1}$, a reliable mass flow of
$12.5$ km\,s$^{-1}$ can cause the onset of an instability of the Kelvin-Helmholtz
type of $m = 1$ mode.  Any none-zero environment magnetic
field, $B_{\rm e}$, increases $M_{\rm A}^{\rm cr}$, slightly for $b = 0.1$,
and more noticeable at $b = 0.5$.  The
observation/detection of the Kelvin-Helmholtz instability of $m = 1$ mode in a
coronal twisted magnetic tube (at $\eta = 0.1$) is generally problematic, but
a twist of $0.4$ might help starting this instability in coronal
X-ray flows, for which the parameter $\eta$ is slightly bigger (of the order
of $0.14$), and the observed flow speeds can be much higher, a few thousand
kilometers per second (see Madjarska \cite{madjarska11}, also
Zhelyazkov \cite{ivan12b} and the discussion therein).

A criterion for the appearance of the Kelvin-Helmholtz instability of kink
waves in untwisted magnetic tubes is the satisfaction of an inequality
suggested by Andries \& Goossens (\cite{andries01}), which in our notation
reads
\[
      M_{\rm A} > 1 + \frac{B_{\rm e}/B_{\rm i}}{\sqrt{\eta}}.
\]
For twisted magnetic tubes, and in particular for isolated ones (with
$B_{\rm e} = 0$), this criterion provides a fairly rough prediction, just
$M_{\rm A} > 1$.  In other words, the mass flows must be super-Apfv\'enic
to trigger an instability -- this seems to be a norm for twisted magnetic
tubes with field-aligned flows (see D\'{i}az et al.\ \cite{diaz11}).  Our
numerical findings of the critical Alfv\'en-Mach numbers also 
agree with that rough criterion -- in both cases ($\eta = 2$ and $0.1$)
the Kelvin-Helmholtz instability occurs when the axial mass flows are
super-Alfv\'enic.

On the other hand, a criterion for the appearance of the
Kelvin-Helmholtz instability in isolated twisted magnetic tubes is the
satisfaction of an inequality suggested by Zaqarashvili et al.\
(\cite{temury10}), which in our notation is
\[
    m M^2_{\rm A} > 1 + 1/\eta.
\]
For the kink mode, i.e., $m = 1$, this inequality gives
$M_{\rm A} > 1.2247$ for $\eta = 2$ and $M_{\rm A} > 3.3166$ for $\eta = 0.1$.
Our numerical findings of the critical Alfv\'en-Mach numbers agree
with this criterion in both cases.

The next steps in studying the Kelvin-Helmholtz instability of kink waves in
twisted magnetic tubes are obviously (i) to investigate the role of the
transverse density inhomogeneity inside the tube and associated with
it resonant wave absorption on the kink waves propagation and their
stability/instability status, and (ii) to consider a field-aligned mass flow
instead of an axial one.

\begin{acknowledgements}
      I.Zh.\ thanks the Sofia University Scientific Fund for support under
grant No.~147-2012.  The work of T.V.Z.\ was supported by the Austrian Fonds zur
F\"orderung der wissenschaftlichen Forschung (project P21197-N16) and by the
European FP7-PEOPLE-2010-IRSES-269299 project-SOLSPANET.  The authors thank
the anonymous referee for his/her constructive comments and suggestions, which led
to significant improvements in the paper.
\end{acknowledgements}

\end{document}